\documentclass[aps,prd,superscriptaddress,showpacs]{revtex4}
\usepackage{graphicx}
\usepackage{epstopdf}
\usepackage{amsmath}
\usepackage{amsfonts}
\usepackage{amssymb}
\usepackage{latexsym}
\usepackage{verbatim}

\begin{document}

\title{Aspects of Lorentz-Poincar\'e-symmetry violating physics in a supersymmetric scenario}
\author{Patricio Gaete}
\email{patricio.gaete@usm.cl}
\affiliation{Departmento de F\'{\i}sica and Centro Cient\'{\i}fico-Tecnol\'ogico de
Valpara\'{\i}so-CCTVal, Universidad T\'{e}cnica Federico Santa Mar\'{\i}a, Valpara\'{\i}so, Chile}
\author{J. A. Helay\"el-Neto}
\email{helayel@cbpf.br}
\affiliation{Centro Brasileiro de Pesquisas F\'\i sicas, Rio de Janeiro, RJ, Brasil}
\author{Alessandro D. A. M. Spallicci }
\email{spallicci@cnrs-orleans, http://wwwperso.lpc2e.cnrs.fr/~spallicci/}
\affiliation{Observatoire des Sciences de l'Univers en r\'egion Centre (OSUC) UMS 3116, Universit\'e d'Orl\'eans, 1A rue de la F\'erollerie, 45071 Orl\'eans, France; P\^ole de Physique, Collegium Sciences et Techniques (CoST), Universit\'e d'Orl\'eans, Rue de Chartres, 45100 Orl\'{e}ans, France; Laboratoire de Physique et Chimie de l'Environnement et de l'Espace (LPC2E) UMR 7328; Centre National de la Recherche Scientifique (CNRS), Campus CNRS, 3A Avenue de la Recherche Scientifique, 45071 Orl\'eans, France; Instituto de F\'{i}sica, Departamento de F\'{i}sica Te\'{o}rica, Universidade do Estado do Rio de Janeiro (UERJ), Rua S\~ao Francisco Xavier 524, 20550-013 Maracan\~a, Rio de Janeiro, Brasil}
\date{\today }

\begin{abstract}
We study a Lorentz invariance violating extension for the pure photonic sector of the standard model in a supersymmetric scenario. We identify a number of independent background fermion condensates. An effective photonic action is proposed which is induced by the SUSY background fermion condensates. The physical consequences leading to direct measurable effects over the screening and confining properties are considered. In the specific case of the Carroll-Field-Jackiw, we pay special attention in analyzing the dispersion relations to derive the photon and photino masses in terms of the supersymmetric background parameters. In connection with astrophysical aspects of Lorentz-symmetry violation, we  discuss the time delay between electromagnetic waves of different frequencies as a consequence of the appearance of the massive photon. We also point out that, in the scenario we propose to accommodate Lorentz-symmetry violation in presence of a supersymmetric background, there is room for a sort of fermionic Primakoff effect in which a photino-photon conversion may be induced by the fermionic sector of the supersymmetric background.
\end{abstract}

\maketitle

\section{Introduction}

It is well known that Lorentz invariance is one of the building blocks of quantum field theory (QFT), which is an exact symmetry implemented by the Standard-Model of Particle Physics (SM) \cite{Colladay97}. This theory relies on (global) Lorentz symmetry and provides a tremendously successful description of nowadays known phenomena. Nevertheless, the observation of Lorentz-Poincar\'e-symmetry violation (LSV) would give a positive signature for the existence of a new and unconventional physics. In fact, the possibility that Lorentz and CPT symmetries be spontaneously broken at very fundamental level, such as in the context of string theories, has driven a very intensive activity. Besides, the necessity of a new scenario has been suggested to overcome theoretical difficulties in the quantum gravity research \cite{Amelino,Urrutia,Piran,Liberati} and to give support to high-precision experimental tests searching for small LSV \cite{Bluhm,Lane,Feng}.  

However, because Lorentz symmetry is at the core of quantum field theory, it is not easy to formulate such a new theoretical structure without a critical revision on its foundations supporting those experiments searching for small LSV corresponding to the low-energy limit of such a yet unknown ultimate theory valid at the Planck scale \cite{Colladay01,Lehnert01}. We mention in passing that, at high energies, supersymmetry (SUSY) is also generally required by theories which attempt to provide an ultimate theory, like string theories. Meanwhile, as our low-energy theories are relativistic QFTs, it is interesting to examine possible new phenomena in this context that could produce deviations from exact Lorentz invariance. Among these researches, the most studied framework is the standard model extension (SME), consisting of the minimal SM plus small Lorentz (and CPT) violating terms, which has provided a theoretical structure for many experiments of Lorentz and CPT violations \cite{pageKostelec}.  

With these observations in mind, in previous studies \cite{susy1,susy2,susy3,Bonetti:2017toa,Bonetti:2016vrq,Helayel-Neto:2019har,Spallicci:2020diu}, we have explored the appealing idea  that the breaking of Lorentz symmetry takes place in a more fundamental level (high-energy scale) and its physical consequences. In this perspective, as soon as this fundamental physics is at work, SUSY might be exact or could be broken in a scale close to the physics of this primary physics. In other words, our idea is that, in a high energy regime, LSV should not be disconnected from SUSY. By following this viewpoint, the scenario for the LSV is that dominated by SUSY or, at least, is influenced by an eventually broken SUSY. As a consequence, SUSY is assumed to be present from the very beginning of our proposal. This then implies that LSV must be originated from some SUSY multiplet. Incidentally, it is of interest notice that this is the main idea of our development. This allows us to find a number of fermion condensates that characterize the background responsible for the LSV, as we will show in this paper.

The line of investigation reported in the present contribution, as already pointed out in the previous paragraphs, contemplate a framework in which (frozen) SUSY degrees of freedom set up the anisotropic background that incorporates deviations from the Lorentz symmetry. This scenario opens up the possibility of connecting the results presented here with  important issues of Astroparticles and Astrophysics. One of these issues is the discussion of the massive photino, whose mass is a by-product of LSV,  as a possible particle to be among the dark matter constituents.  Also, in one of the models we consider here, LSV yields the appearance of  a massive photon in the spectrum. From the (massive) photonic dispersion relation, a physical consequence that we discuss is the computation of the time delay between electromagnetic waves of different frequencies. This time delay is expressible in terms of the SUSY-originated parameter that characterizes the LSV in the specific model. These points shall be duly discussed in the sequel. 

In this perspective, we also point out that these condensates characterize the new vacuum of the theory with striking consequences over the different phases of the electromagnetic sector of an extended standard model. Our interest here is to examine the physical implications of these condensates on physical observables, where the notion of confinement is of fundamental phenomenological significance. It is worth recalling at this point that one of the long-standing issues in gauge theories is a quantitative description of confinement. Nevertheless, phenomenological models still represent a key tool for understanding confinement physics, and can be considered as effective theories of QCD. This has helped us to gain insights over confinement in different theories \cite{Gaete1,Gaete2,Gaete3}.

Inspired by the above observations, in this article we present a self-contained discussion of the implications of  Lorentz-symmetry violating in a supersymmetric scenario based on our previous studies \cite{susy1,susy2,susy3}. 
Our work is organized according to the following outline: in Section $2$, some of the relevant aspects of our development are  reviewed for completeness, and to make the fundamental ideas available to a wider audience. Specifically, in Subsection $2.1$, we briefly reconsider the calculation of the interaction energy between static point-like sources for the gauge sector of a $N=1$ $D=4$ supersymmetric model, which is also known as $B \wedge F$ model \cite{Euro,Aurilia}. This would not only provide the theoretical setup for our subsequent work, but also fix the notation. In Subsection $2.2$ we consider the spectrum of the minimal supersymmetric extension of the Carroll-Field-Jackiw model \cite{CFJ} with a topological Chern-Simons-like Lorentz-symmetry violating term. In such a case, we identify a number of independent background fermion condensates and obtain a photonic effective action. This effective action allows us to discuss the interaction energy induced by the SUSY background fermion condensates. Interestingly enough, the static potential profile contains a linear term, leading to confinement of static charges. Next, in Subsection $2.3$, we assume that LSV takes place in an environment dominated by SUSY. In this case, Lorentz symmetry is violated in the photon sector by an CPT-even $k_{F}$ term \cite{Colladay98}. An effective photonic action is obtained which originates from the SUSY background fermion condensates. Again, following our earlier procedure, the interparticle potential contains a linear term leading to confinement. We shall proceed, in Subsection $2.4$, to study the dimensional reduction for a $(1+3)$-dimensional Lorentz violation Lagrangian in the matter and gauge sectors in a supersymmetric scenario. We focus our attention on a $(1+2)$- dimensional space-time, so that planar phenomena may be considered. We then obtain an effective model for photons induced by the effects of SUSY in our framework with LIV. As in the previous cases, this model is discussed in connection with the calculation of the interaction energy between charged particles. Again, as in the preceding cases, we find a confining potential. It should be further noted that although planar physics must directly interest condensed matter phenomena and those, in principle, do not immediately concern Lorentz symmetry, it is well known that non-relativistic effects derived from Lorentz-invariant models are relevant for condensed matter physics. Finally, some concluding remarks are made in Section $3$.

In our conventions the signature of the metric is $(+1,-1,-1,-1)$ and $(+1,-1,-1)$.

\section{Lorentz-symmetry violation and supersymmetry}

As mentioned above, the purpose of this Section is to further elaborate on the physical content on Lorentz-symmetry violating physics in a supersymmetric scenario.

\subsection{Gauge sector of a supersymmetric model: an example}

As stated in the introduction, we shall now reassess the interaction energy between static point-like sources for the gauge sector of a supersymmetric model. To do this, we will use the gauge-invariant but path-dependent variables formalism, which is an alternative to the Wilson loop approach. According this formalism, the interaction energy between two charges static is obtained once a judicious identification of the physical degrees of freedom is made \cite{Gaete1}. This can be done by computing the expectation value of the energy operator $H$ in the physical state $\left| \Phi  \right\rangle$ describing the sources, which we will denote by ${\left\langle H \right\rangle _\Phi }$.

However, before going to the derivation of the interaction energy, we will describe very briefly the model under consideration. The starting-point of our present discussion is provided by the Lagrangian density:
\begin{equation}
{\cal L} =  - \frac{1}{4}F_{\mu \nu }^2\left( A \right) + \frac{1}{{12}}H_{\mu \nu \rho }^2\left( B \right) 
+ \frac{m}{{24}}{\varepsilon ^{\mu \nu \rho \sigma }}{B_{\mu \nu }}{\partial _{[\rho }}{A_{\sigma ]}},  \label{susy05}
\end{equation}

It is convenient to rewrite this equation in the alternative form 
\begin{equation}
{{\cal L}} =  - \frac{1}{4}F_{\mu \nu }^2 - \frac{1}{2}{\tilde H_\sigma }{\tilde H^\sigma } 
- \frac{m}{6}{\tilde H^\sigma }{A_\sigma }, \label{susy10}
\end{equation}
where we have made use of ${\tilde H^\mu } = 
{\raise0.5ex\hbox{$\scriptstyle 1$}\kern-0.1em/\kern-0.15em\lower0.25ex\hbox{$\scriptstyle 2$}}{\varepsilon ^{\mu \nu \lambda \rho }}{\partial _\nu }{B_{\lambda \rho }}$.

It should, however, be noted here that by eliminating the dual-field $H^{\sigma}$ induces an effective theory for the $A_{\mu}$ field \cite{Aurilia}. Nevertheless, to eliminate this dual-field care must be taken, for it satisfies the constraint ${\partial _\mu }{\tilde H^\mu } = 0$ (Bianchi identity). Accordingly, to take into account the constraint, we shall introduce a Lagrange multiplier $\chi$. Thus, we may now rewrite the foregoing equation in the form 
\begin{equation}
{\cal L} = - \frac{1}{4}F_{\mu \nu }^2 - \frac{1}{2}{\tilde H_\sigma }{\tilde H^\sigma }
 - \frac{m}{6}{\tilde H^\sigma }{A_\sigma } + \chi {\partial _\sigma }{\tilde H^\sigma }. \label{susy15}
\end{equation}
By introducing ${Z_\sigma } \equiv {A_\sigma } + \frac{6}{m}{\partial _\sigma }\chi$, with $  {Z_{\mu \nu }} 
= {F_{\mu \nu }}$, we get
\begin{equation}
{\cal L} =  - \frac{1}{4}Z_{\mu \nu }^2 - \frac{1}{2}{\tilde H_\sigma }{\tilde H^\sigma } 
- \frac{m}{6}{\tilde H^\sigma }{Z_\sigma }.\label{susy20} 
\end{equation}
By a further definition of the fields, ${W_\sigma } \equiv {\tilde H_\sigma } + \frac{m}{6}{Z_\sigma }$, we find that the Lagrangian density (\ref{susy05}) reduces to
\begin{equation}
{\cal L} =  - \frac{1}{4}Z_{\mu \nu }^2 + \frac{1}{2}{\mu ^2}Z_\mu ^2, \label{susy25}
\end{equation}
with ${\mu ^2} \equiv {\raise0.5ex\hbox{$\scriptstyle {{m^2}}$}
\kern-0.1em/\kern-0.15em
\lower0.25ex\hbox{$\scriptstyle {36}$}}$. We immediately see that the Lagrangian density (\ref{susy25}) exhibits a de Broglie-Proca-type mass term.

Having established the new equivalent Lagrangian, we can now compute 
the interaction energy between external probe sources for the model under consideration. 
To do this, however, our first undertaking is to restore the gauge invariance in (\ref{susy25}). 
To this end we shall use the Hamiltonian formalism for constrained systems along the lines of
Ref. \cite{Gaete1}. The canonically conjugate are $\Pi ^0=0$ and 
$\Pi ^i = - F^{0i}$. The canonical Hamiltonian is then,
\begin{equation}
{H_C} = \int {{d^3}x} \left\{ { - {A_0}\left\{ {{\partial _i}{\Pi ^i} + \frac{{{\mu ^2}}}{2}{A^0}} \right\} - \frac{1}{2}{\Pi _i}{\Pi ^i} + \frac{1}{4}{F_{ij}}{F^{ij}} - \frac{1}{2}{A_i}{\mu ^2}{A^i}} \right\}.
 \label{susy30}
\end{equation}
Next, the primary constraint, $\Pi ^0=0$, must be satisfied for all times. Accordingly, we obtain the
following secondary constraint
\begin{equation}
\Gamma \left( x \right) \equiv \partial _i \Pi ^i  + \frac{\mu^2}{2} A^0  = 0. \label{susy35}
\end{equation}
From the preceding discussion we easily verify the second class nature of the constraints, as expected for a theory with an explicit mass term which breaks the gauge invariance. In view of this situation, as was explained in Ref. \cite{Gaete1}, we enlarge the original phase space by
introducing a canonical pair of fields $\theta$ and $ \Pi _\theta $. Through this procedure,
a new set of constraints can be defined in this extended space:
\begin{equation}
\Lambda _1  \equiv \Pi _0  + \frac{\mu^2}{2} \theta, \label{susy40a}
\end{equation}
and
\begin{equation}
\Lambda _2  \equiv \Gamma  + \Pi _\theta. \label{susy40b}
\end{equation}
It is a simple matter to verify that the new constraints are first class which, then, restores
the gauge symmetry of the theory under consideration. Thus, we finally obtain a new effective
Lagrangian, which is given by
\begin{equation}
\mathcal{ L} = - \frac{1}{4}{F_{\mu \nu }}\left( {1 + \frac{{{\mu ^2}}}{\Delta }} \right){F^{\mu \nu }}.
\label{susy45}
\end{equation}
In this last line we have integrated out the $\theta$ field, and $\Delta  \equiv {\partial _\mu }{\partial ^\mu }$.

Once this is done, the canonical quantization of this theory from the Hamiltonian point of view is straightforward. The canonical momenta read ${\Pi ^\mu } =  - \left( {1 + \frac{{{\mu ^2}}}{\Delta }} \right){F^{0\mu }}$, which
produces the usual primary constraint $\Pi ^0  = 0$ and ${\Pi ^i} =  - \left( {1 + \frac{{{\mu ^2}}}{\Delta }} \right){F^{0i}}$. The canonical Hamiltonian is thus
\begin{equation}
{H_C} = \int {{d^3}x} \left\{ { - {A_0}{\partial _i}{\Pi ^i} - \frac{1}{2}{\Pi _i}{{\left( {1 + \frac{{{\mu ^2}}}{\Delta }} \right)}^{ - 1}}{\Pi ^i} + \frac{1}{4}{F_{ij}}\left( {1 + \frac{{{\mu ^2}}}{\Delta }} \right){F^{ij}}} \right\}. \label{susy50}
\end{equation}
Hence the consistency condition ${\dot \Pi _0} = 0$ leads to the secondary constraint $\Gamma_1 \left( x \right) \equiv \partial _i \Pi ^i=0$. It can be easily seen that the stability of this constraint does not generate
further constraints. In accordance with the Dirac method, the extended Hamiltonian that
generates translations in time then reads $H = H_C + \int {d^3
}x\left( {c_0 \left( x \right)\Pi _0 \left( x \right) + c_1 \left(
x\right)\Gamma _1 \left( x \right)} \right)$, where $c_0 \left(
x\right)$ and $c_1 \left( x \right)$ are arbitrary Lagrange
multipliers. However it is easily seen that $\dot{A}_0 \left( x \right)= \left[ {A_0 \left( x \right),H} \right] = c_0 \left( x \right)$, which is an arbitrary function. Since $ \Pi^0 = 0$ always, neither $A^0 $ nor $ \Pi^0 $ are of interest in describing the system and may be eliminated from the theory. In summary then, the Hamiltonian simplifies to 
\begin{equation}
{H} = \int {{d^3}x} \left\{ { c(x){\partial _i}{\Pi ^i} - \frac{1}{2}{\Pi _i}{{\left( {1 + \frac{{{\mu ^2}}}{\Delta }} \right)}^{ - 1}}{\Pi ^i} + \frac{1}{4}{F_{ij}}\left( {1 + \frac{{{\mu ^2}}}{\Delta }} \right){F^{ij}}} \right\}              , \label{susy55}
\end{equation}
where $c(x) = c_1 (x) - A_0 (x)$.

The quantization of the theory requires the removal of nonphysical variables, which is done by imposing a gauge condition such that the full set of constraints become second class. A particularly convenient choice is found to be
\begin{equation}
\Gamma _2 \left( x \right) \equiv \int\limits_{C_{\xi x} } {dz^\nu }
A_\nu \left( z \right) \equiv \int\limits_0^1 {d\lambda x^i } A_i
\left( {\lambda x} \right) = 0,     \label{susy60}
\end{equation}
where  $\lambda$ $(0\leq \lambda\leq1)$ is the parameter describing
the spacelike straight path $ x^i = \xi ^i  + \lambda \left( {x -
\xi } \right)^i $, and $ \xi $ is a fixed point (reference point).
There is no essential loss of generality if we restrict our
considerations to $ \xi ^i=0 $. The choice (\ref{susy60}) leads to
the Poincar\'e gauge \cite{Gaete1}. We thus find that the nontrivial Dirac bracket for the canonical
variables is given by
\begin{equation}
\left\{ {A_i \left( x \right),\Pi ^j \left( y \right)} \right\}^ *
=\delta{ _i^j} \delta ^{\left( 3 \right)} \left( {x - y} \right) -
\partial _i^x \int\limits_0^1 {d\lambda x^j } \delta ^{\left( 3
\right)} \left( {\lambda x - y} \right). \label{susy65}
\end{equation}

We now proceed to determine the interaction energy for the model under consideration. As mentioned above, 
to do that we need to compute the expectation value of the energy operator $H$ in the physical state
$|\Phi\rangle$. Now we recall that the physical state $|\Phi\rangle$ can be written as \cite{Dirac}
\begin{equation}
\left| \Phi  \right\rangle  \equiv \left| {\overline \Psi  \left(
\bf y \right)\Psi \left( {\bf y}\prime \right)} \right\rangle
= \overline \psi \left( \bf y \right)\exp \left(
{iq\int\limits_{{\bf y}\prime}^{\bf y} {dz^i } A_i \left( z \right)}
\right)\psi \left({\bf y}\prime \right)\left| 0 \right\rangle,
\label{susy70}
\end{equation}
where the line integral is along a space-like path on a fixed time slice, and $\left| 0 \right\rangle$ is the physical vacuum state. Also it is important to point out that the above expression clearly shows that each of the states $(|\Phi\rangle)$ represents a fermion-antifermion pair surrounded by a cloud of gauge fields to maintain gauge invariance.

Making use of the foregoing Hamiltonian structure, we observe that
\begin{equation}
 {\Pi _i}\left( x \right)\left| {\overline \Psi  \left( {\bf y} \right)\Psi \left( {{{\bf y}^ \prime }} \right)} \right\rangle  = \overline \Psi  \left( {\bf y} \right)\Psi \left( {{{\bf y}^ \prime }} \right){\Pi _i}\left( x \right)\left| 0 \right\rangle  + q\int_{\bf y}^{{{\bf y}^ \prime }} {d{z_i}{\delta ^{\left( 3 \right)}}} \left( {{\bf z} - {\bf x}} \right)\left| \Phi  \right\rangle. \label{susy75}
 \end{equation}
 
Having made this observation and and since the fermions are taken to be infinitely massive (static) we can
substitute $\Delta$ by $-\nabla^{2}$ in Eq. (\ref{susy55}). Therefore, the expectation value $\left\langle H \right\rangle _\Phi$ becomes
\begin{equation}
\left\langle H \right\rangle _\Phi   = \left\langle H \right\rangle
_0 + \left\langle H \right\rangle _\Phi ^{\left( 1 \right)},
\label{susy80}
\end{equation}
where $\left\langle H \right\rangle _0  = \left\langle 0
\right|H\left| 0 \right\rangle$, and the $\left\langle H \right\rangle
 _\Phi ^{\left( 1 \right)}$ term is given by
\begin{equation}
\left\langle H \right\rangle _\Phi ^{\left( {1} \right)}  =  - \frac{q^2}{2}
\int {d^3 x} \int_{\bf y}^
{{\bf y}^ \prime  } {dz_i^ \prime  } \delta ^{\left( 3 \right)} \left( {{\bf x} -
{\bf z}^ \prime  } \right)\left( {1 - \frac{{\mu^2 }}{{\nabla ^2 }}} \right)_x^{ - 1}
\int_{\bf y}^{{\bf y}^ \prime  } {dz^i }
\delta ^{\left( 3 \right)} \left( {{\bf x} - {\bf z}} \right), \label{susy85}
\end{equation}
where the integrals over $z^{i}$ and $z_{i}^{\prime}$ are zero except on the contour of integration.

One can now further observe that the $\left\langle H \right\rangle _\Phi ^{\left( {1} \right)}$ term may look peculiar, but it is nothing but the usual Yukawa interaction plus self-energy terms. As we have already noticed in \cite{GaeteWot}, equation  (\ref{susy85}) can be brought to the form
\begin{equation}
\left\langle H \right\rangle _\Phi ^{\left( 1 \right)} = \frac{{{q^2}}}{2}\int_{\bf y}^{{{\bf y}^ \prime }} {dz_i^ * } \partial _i^{{z^ * }}\int_{\bf y}^{{{\bf y}^ \prime }} {d{z^i}} \partial _z^iG\left( {{{\bf z}^ \prime },{\bf z}} \right), \label{susy86}
\end{equation}
where $G$ is the Green function $G\left( {{{\bf z}^ \prime },{\bf z}} \right) = \frac{1}{{4\pi }}\frac{{{e^{ - \mu |{{\bf z}^ \prime } - {\bf z}|}}}}{{|{{\bf z}^ \prime } - {\bf z}|}}$. Making use of the foregoing results, we find that the potential for two opposite charges located at ${\bf y}$ and ${\bf y}^{\prime}$ takes the form
\begin{equation}
V = -\frac{{q^2 }}{{4\pi }}\frac{{e^{ - \mu|{\bf y} - {\bf y}^ \prime  |} }}
{{|{\bf y} - {\bf y}^ \prime  |}}, \label{susy90}
\end{equation}
after subtracting the self-energy terms.
 
It is appropriate to observe here that there is an alternative but equivalent way of obtaining the result (\ref{susy90}). To this end we consider the potential \cite{Gaete1}
\begin{equation}
V \equiv q\left( {{\cal A}_0 \left( {\bf 0} \right) - {\cal A}_0
\left( {\bf y} \right)} \right), \label{susy110}
\end{equation}
where the physical scalar potential is given by
\begin{equation}
{\cal A}_0 \left( {x^0 ,{\bf x}} \right) = \int_0^1 {d\lambda } x^i
E_i \left( {\lambda {\bf x}} \right), \label{susy115}
\end{equation}
with $i=1,2,3$. This equation follows from the vector gauge-invariant field expression \cite{Gaete1}
\begin{equation}
{\cal A}_\mu  \left( x \right) \equiv A_\mu  \left( x \right) +
\partial _\mu  \left( { - \int_\xi ^x {dz^\mu  } A_\mu  \left( z
\right)} \right), \label{susy120}
\end{equation}
where, as in Eq.(\ref{susy70}), the line integral is along a spacelike path from the point $\xi$ to $x$, on a fixed slice time. It is of interest also to notice that the gauge-invariant variables (\ref{susy120}) commute with the sole first constraint (Gauss law), corroborating  in this way that these fields are physical variables \cite{Dirac}.

With this in hand, we recall that Gauss law for the present theory is given by $\partial _i \Pi ^i  = J^0$, where we have included the external current $J^0$ to represent the presence of two opposite charges. For $J^0 \left( {t,{\bf x}} \right) = q\delta ^{\left( 3 \right)} \left( {\bf x} \right)$, we readily deduce that
\begin{equation}
{\cal A}_0 \left( {\bf x} \right) = - qG({\bf x}),  \label{susy125}
\end{equation}
after substraction of self-energy terms and $G({\bf x}) = \frac{{{e^{ - \mu |{\bf x}|}}}}{{4\pi |{\bf x}|}}$ is the previous Green function. This, together with Eq.(\ref{susy110}), then, yields 
\begin{equation}
V = -\frac{{q^2 }}{{4\pi }}\frac{{e^{ - \mu|{\bf y} - {\bf y}^ \prime  |} }}
{{|{\bf y} - {\bf y}^ \prime  |}}, \label{susy130}
\end{equation}
for a pair of point-like opposite charges q located at ${\bf y}$ and ${\bf y^{\prime}}$. 
Finally, we call attention to the fact that a correct identification of physical degrees of
freedom is a crucial feature for understanding the physics hidden in gauge theories. In this way one encounters that, once the identification is made, the computation of the potential is carried out by means of Gauss law.

\subsection{Supersymmetric extension of the Carroll-Field-Jackiw model for electrodynamics I}

In what follows, we address the possibility to realize a specific model of LSV in a scenario dominated by SUSY, as originally developed in our Ref. \cite{susy1} which we follow closely.

To do this we start with the modified supersymmetric Abelian gauge model proposed in \cite{Belich:2003fa}, which is a superfield version of the Carroll-Field-Jackiw Electrodynamics \cite{CFJ} with a background superfield that realizes the Lorentz-symmetry breaking. Interestingly, it is observed that this model preserves supersymmetry at the action level, while the Lorentz symmetry is violated in the sense of particle transformations. By following a covariant superspace-superfield formulation \cite{BaetaScarpelli:2003kx}, we have:
\begin{equation}
S_{LSV}=\int d^{4}xd^{2}\theta d^{2}\bar{\theta}\left\{ W^{a}(D_{a}V)S+\overline{W}
_{\dot{a}}(\overline{D}^{\dot{a}}V)\overline{S}\right\} ,  \label{susy135}
\end{equation}
where the superfields $W_{a}$, $V$, $S$ and the susy-covariant derivatives, $
\ D_{a}$, $\overline{D}_{\dot{a}}$, are given by:
\begin{eqnarray}
D_{a} &=&\frac{\partial }{\partial \theta ^{a}}+i{\sigma ^{\mu }}_{a\dot{a}}
\bar{\theta}^{\dot{a}}\partial _{\mu } \\
\overline{D}_{\dot{a}} &=&-\frac{\partial }{\partial \bar{\theta}^{\dot{a}}}
-i\theta ^{a}{\sigma ^{\mu }}_{a\dot{a}}\partial _{\mu }; \label{susy140}
\end{eqnarray}
whereas the field-strength superfield, $W_a$, reads
\begin{equation}
W_{a}(x,\theta ,\bar{\theta})=-\frac{1}{4}\overline{D}^{2}D_{a}V. \label{susy145}
\end{equation}
Another interesting observation is that the action (\ref{susy135}) is gauge invariant up to surface terms. In fact, the Bianchi identities $D^a W_a  = \bar D_{\dot a} \bar W^{\dot a}  = 0$ and the constraints on $S$ and $\bar S$ (given in the sequel) ensure gauge invariance of our Lorentz-symmetry violating action. We observe that, $W_a$, can be $\theta $-expanded in the form: 
\begin{eqnarray}
{W_a}\left( {x,\theta ,\bar \theta } \right) &=& {\lambda _a}\left( x \right) + i{\theta ^b}\sigma _{b\dot a}^\mu {{\bar \theta }^{\dot a}}{\partial _\mu }{\lambda _a}\left( x \right) - \frac{1}{4}{{\bar \theta }^2}{\theta ^2}{\lambda _a}\left( x \right) + 2{\theta _a}D\left( x \right) - i{\theta ^2}{{\bar \theta }^{\dot a}}{\sigma ^\mu }_{a\dot a}{\partial _\mu }D\left( x \right)    \nonumber\\
 &+& {\sigma ^{\mu \nu }}{_a^b}{\theta _b}{F_{\mu \nu }}\left( x \right) - \frac{i}{2}{\sigma ^{\mu \nu }}{_a^b}{\sigma ^\alpha }_{b\dot a}{\theta ^2}{{\bar \theta }^{\dot a}}{\partial _\alpha }{F_{\mu \nu }}\left( x \right) - i{\sigma ^\mu }_{a\dot a}{\partial _\mu }{{\bar \lambda }^{\dot a}}\left( x \right){\theta ^2}, \label{susy150}
\end{eqnarray}
and $V=V^{\dagger }$ is the so-called gauge-potential superfield, which is a real scalar. We further note that, as is usually done to perform component-field calculations, we adopt the Wess-Zumino gauge:
\begin{equation}
\text{ }V_{WZ}=\theta \sigma ^{\mu }\bar{\theta}A_{\mu }(x)+\theta ^{2}\bar{
\theta}\overline{\lambda }\left( x\right) +\bar{\theta}^{2}\theta \lambda
(x)+\theta ^{2}\bar{\theta}^{2}D. \label{susy155}
\end{equation}
Next, the background superfield, $S$, is chosen to be a chiral supermultiplet. Such a constraint restricts the
highest spin component of the background to be an $s$ $=$ $\frac{1}{2}$
component-field. Also, according to the action of eq.(\ref{susy135}), one should notice that $S$ happens to be dimensionless. As a physical propagating superfield, its mass dimension would be equal to $1$. The $\theta$-expansion for the background superfield $S$ then reads:  
\begin{equation}
\overline{D}_{\dot{a}}S\left( x\right) =0, \label{susy160}
\end{equation}
and 
\begin{equation}
S\left( x \right) = s\left( x \right) + i\theta {\sigma ^\mu }\bar \theta {\partial _\mu }s\left( x \right) - \frac{1}{4}{\bar \theta ^2}{\theta ^2}s\left( x \right) + \sqrt 2 \theta \psi \left( x \right) + \frac{i}{{\sqrt 2 }}{\theta ^2}\bar \theta {\bar \sigma _\mu }{\partial _\mu }\psi \left( x \right) + {\theta ^2}F\left( x \right). \label{susy165}
\end{equation}

The component-wise counterpart for the action (\ref{susy135}) is given by the expression (after the supersymmetric Maxwell action is also included): 
\begin{eqnarray}
S_{comp.} &=&\int {d^{4}x}\Biggl[-\left\{ {\frac{1}{4}+\frac{{\left( {
s+s^{\ast }}\right) }}{2}}\right\} F_{\mu \nu }F^{\mu \nu }+\frac{i}{2}
\partial _{\mu }\left( {s-s^{\ast }}\right) \varepsilon ^{\mu \alpha \beta
\nu }F_{\alpha \beta }A_{\nu }+\left\{ {\frac{1}{2}+4\left( {s+s^{\ast }}
\right) }\right\} D^{2}  \notag \\
&-&\left( {\frac{1}{2}-2s}\right) i\lambda \sigma ^{\mu }\partial _{\mu }
\bar{\lambda}-\left( {\frac{1}{2}-2s^{\ast }}\right) i\bar{\lambda}\bar{
\sigma}^{\mu }\partial _{\mu }\lambda -\sqrt{2}\lambda \sigma ^{\mu \nu
}\psi F_{\mu \nu }+\sqrt{2}\bar{\lambda}\bar{\sigma}^{\mu \nu }\bar{\psi}
F_{\mu \nu }  \notag \\
&+&\lambda \lambda F+\bar{\lambda}\bar{\lambda}F^{\ast }-2\sqrt{2}\lambda
\psi D-2\sqrt{2}\bar{\lambda}\bar{\psi}D\Biggr]\ .  \label{susy170}
\end{eqnarray}
By suitably choosing $s$, such that $s+s^{\ast }=0$ , it is the imaginary part of $s$ the
responsible for the appearance of the vector $v_{\mu }$ of the
Carroll-Field-Jackiw term; $s-s^{\ast }=-(i/2)v_{\mu }x^{\mu }$. D is fixed by its algebraic field
equation, $D=\sqrt{2}(\lambda \psi +\bar{\lambda}\bar{\psi})$. Making
use of Fierz rearrangements in all the $4$-fermion terms and rewriting the action in terms of Majorana 4-component spinors, we arrive at 
\begin{eqnarray}
{S_{comp}} = \int {{d^4}x} \left[ { - \frac{1}{4}{F_{\mu \nu }}{F^{\mu \nu }} + \frac{1}{4}{v_\mu }{\varepsilon ^{\mu \alpha \beta \nu }}{F_{\alpha \beta }}{A_\nu } - \frac{i}{2}\bar \Lambda {{\bar \gamma }^\mu }{\partial _\mu }\Lambda  + \left( {{\mathop{\rm Re}\nolimits} \left( F \right) + \frac{1}{4}\bar \Psi \Psi } \right)\bar \Lambda \Lambda } \right]  \nonumber\\
 + \int {{d^4}x} \left[ { - i \left( {{\mathop{\rm Im}\nolimits} \left( F \right) + \frac{i}{4}\bar \Psi {\gamma _5}\Psi } \right)\bar \Lambda {\gamma _5}\Lambda  - \frac{1}{4}\left( {{v_\mu } + \bar \Psi {\gamma _\mu }{\gamma _5}\Psi } \right)\left( {\bar \Lambda {\gamma ^\mu }{\gamma _5}\Lambda } \right) + \sqrt 2 \bar \Lambda {\Sigma ^{\mu \nu }}{\gamma _5}\Psi {F_{\mu \nu }}} \right], \label{susy175}
\end{eqnarray}
with the Majorana fermions, $\Lambda$ (the gaugino) and $\Psi$ (the background fermion), given by:
\begin{equation}
\Lambda \equiv \left( 
\begin{array}{c}
\lambda _{\alpha } \\ 
\bar{\lambda}_{\dot{\alpha}}
\end{array}
\right),\, \,\,\Psi \equiv \left( 
\begin{array}{c}
\psi _{\alpha } \\ 
\bar{\psi}_{\dot{\alpha}}
\end{array}
\right), 
\end{equation} \label{susy180}
and
\begin{equation}
\Sigma ^{\mu \nu }  \equiv \frac{i}{4}\left[ {\gamma ^\mu  ,\gamma ^\nu  } \right]. \label{susy185}
\end{equation}

This leads to the field equations: 
\begin{equation}
{\partial _\mu }{F^{\mu \nu }} + {v_\mu }{\tilde F^{\mu \nu }} + 2\sqrt 2 \,\bar \Psi \,{\Sigma ^{\mu \nu }}{\gamma _5}\left( {{\partial _\mu }\Lambda } \right) = 0. \label{susy185b}
\end{equation}

In addition, we also have
\begin{equation}
i{\gamma ^\mu }{\partial _\mu }\Lambda  - 2{M_1}\Lambda  + 2i{M_2}\,{\gamma _5}\,\Lambda  + \frac{1}{2}{R_\mu }{\gamma ^\mu }{\gamma _5}\,\Lambda  = \sqrt 2\, {\Sigma ^{\mu \nu }}{\gamma _5}\Psi {F_{\mu \nu }}, \label{susy185d}
\end{equation}
where $M_{1}= \left( {{\mathop{\rm Re}\nolimits} \left( F \right) + \frac{1}{4}\bar \Psi \Psi } \right)$, $M_{2}= \left( {{\mathop{\rm Im}\nolimits} \left( F \right) + \frac{i}{4}\bar \Psi {\gamma _5}\Psi } \right)$, and $R_{\mu}= \left( {{v_\mu } + \bar \Psi {\gamma _\mu }{\gamma _5}\Psi } \right)$.

Let us also mention here that, by iterating equation (\ref{susy185d}) and taking the Dirac conjugate, we get the Gordon-type decomposition below for the photino current:
\begin{equation}
\left( {p_\mu ^ \prime  + {p_\mu }} \right)\bar \Lambda \left( {{p^ \prime }} \right){\Sigma ^{\mu \kappa }}\Lambda \left( p \right) = \left( {{v_\mu } + \bar \Psi {\gamma _\mu }{\gamma _5}\Psi } \right)\bar \Lambda \left( {{p^ \prime }} \right){\Sigma ^{\mu \kappa }}{\gamma _5}\Lambda \left( p \right), \label{susy185e}
\end{equation}
and
\begin{eqnarray}
\left( {{p^\kappa } - {p^{ \prime \kappa }}} \right)\bar \Lambda \left( {{p^ \prime }} \right)\Lambda \left( p \right) - 2\,i\left( {{p_\mu } + p_\mu ^ \prime } \right)\bar \Lambda \left( {{p^ \prime }} \right){\Sigma ^{\kappa \mu }}\Lambda \left( p \right) + 4\,i\,{M_2}\bar \Lambda \left( {{p^ \prime }} \right){\gamma ^\kappa }{\gamma _5}\Lambda \left( p \right) + \nonumber\\
\left( {{v^\kappa } + \bar \Psi {\gamma ^\kappa }{\gamma ^5}\,\Psi } \right)\bar \Lambda \left( {{p^ \prime }} \right){\gamma _5}\Lambda \left( p \right) =  - i\,2\,\sqrt 2\, \bar \Psi {\gamma _\mu }{\gamma _5}\Lambda \left( p \right){F^{\mu \kappa }}\left( {{p^ \prime }} \right) + 2\,\sqrt 2\, \bar \Psi {\gamma _\mu }\Lambda \left( p \right){{\tilde F}^{\mu \kappa }}\left( {{p^ \prime }} \right). \label{susy185f}
\end{eqnarray}

It must be clear from eq. (\ref{susy175}), how the fermionic background, $\Psi$, and the scalar, $F$, affect the photino sector of SUSY in eq. (\ref{susy175}): they yield mass-type terms for $\Lambda$.
In the same way, we highlight the presence of a new photon-photino term (the very last term of (\ref{susy175})), which appears due exclusively to the presence of the fermionic component of the background.

It is also important to observe that, by inspecting the SUSY transformations of the component fields, the breaking of Lorentz symmetry necessarily implies the appearance of a sort of Goldstino particle.
Since $\partial _{\mu }B$, being non-trivial, yields $\delta \Psi \neq 0$, once the SUSY variation of $\Psi$ reads as follows:
\begin{equation}
\delta \Psi =\partial _{\mu }(A-\gamma _{5}B)\gamma ^{\mu }\varepsilon
+f\varepsilon +g\gamma _{5}\varepsilon , \label{susy190}
\end{equation}
where $A$, $B$, $f$ and $g$ are such that $s=A+iB$, $F=f+ig$. $\varepsilon $ is the four-component
Majorana parameter of the SUSY transformation. This signals the presence of a Goldstone fermion produced as a perturbation around the background, even if $f=g=0$, but with $\partial _\mu  B =  - \frac{1}{4}v_\mu$. So, SUSY is broken together with Lorentz symmetry. Translations are not broken, since $v^{\mu}$ is constant and then no explicit $x^{\mu}$-dependence is present in (\ref{susy175}) through the background components fields ($
\Psi$ and $F$ are also  $x^{\mu}$-independent). Accordingly, Poincar\'e symmetry is actually broken only in the sector of boosts and space rotations.

With the idea of keeping track of the effects of the fermionic background we now consider the dispersion relations.
In other words, to read off the photon and photino dispersion relations in the presence of the complete background responsible for the LSV, namely $ 
\left\{ {A,B,\Psi _\alpha  ,f,g} \right\}$, it is suitable to express the the kinetic part of the Lagrangian, taking account that the background fields are fixed, in the form that is cast below:
\begin{equation}
{\mathcal{L}}=\frac{1}{2}{\Phi ^{t}\mathcal{O}}\Phi =\frac{1}{2}\left( 
\begin{array}{cc}
\bar{\Lambda}_{a} & A_{\mu }
\end{array}
\right) \left( 
\begin{array}{cc}
J^{ab} & L^{a\nu } \\ 
M^{\mu b} & N^{\mu \nu }
\end{array}
\right) \left( 
\begin{array}{c}
\Lambda _{b} \\ 
A_{\nu }
\end{array}
\right) ,  \label{susy195}
\end{equation}
where 
\begin{equation}
J^{ab} =-i(\gamma ^{\mu }\partial _{\mu })^{ab}+(2\text{Re}(F)+\frac{\mu}{2})  \delta^{ab} -i(2\text{Im}(F)+i\frac{\tau}{2})\gamma
_{5}^{ab}-\frac{1}{2}\left( v_\mu+\bar{\Psi}\gamma _{\mu }\gamma _{5}\Psi \right) \left(
\gamma ^{\mu }\gamma _{5}\right) ^{ab}, \label{susy200}
\end{equation}
\begin{equation}
L^{a\nu }  = 2\sqrt 2 \left( {\Sigma ^{\mu \nu } \gamma _5 } \right)^{ab} \Psi \partial _\mu, \label{susy205}
\end{equation}
\begin{equation}
M^{\mu b} =2\sqrt{2}\bar{\Psi}(\Sigma ^{\nu \mu }\gamma _{5})^{ab}\partial_{\nu }, \label{susy210}
\end{equation}
and
\begin{equation}
{N^{\mu \nu }} = {\theta ^{\mu \nu }} - {v_\rho }{\varepsilon ^{\rho \lambda \mu \nu }}{\partial _\lambda } - \frac{1}{\alpha }{w^{\mu \nu }}. \label{susy215}
\end{equation}
As we usually proceed, a gauge-fixing term with parameter $\alpha $ is introduced to ensure invertibility of $N$. In case we wished to explicitly read off the photon-photino propagators, we would have to compute ${\cal O}^ {-1}$. Since $J$ is invertible, ${\cal O}$ becomes non-singular whenever $N$ is also invertible. In (\ref{susy200})-(\ref{susy215}), we have defined 3 background fermion condensates:
\begin{equation}
\mu  \equiv \bar \Psi \Psi, \label{susy220}
\end{equation}
\begin{equation}
\tau  \equiv \bar \Psi \gamma _5 \Psi, \label{susy225}
\end{equation}
\begin{equation}
C^\mu   \equiv \bar \Psi \gamma ^\mu  \gamma _5 \Psi. \label{susy230}
\end{equation}

Since $\Psi $ is a Majorana spinor, we can ensure that $\mu $ is real, $\tau $ is purely imaginary and $B^{\mu }$ is a pseudo-vector with real components.
Upon some Fierzings and by considering that the Majorana $\Psi $-components are Grassmann-valued, we can readily show that 
\begin{equation}
\mu ^{2}=-\tau ^{2}=\frac{1}{4}C_{\mu }C^{\mu }.  \label{susy235}
\end{equation}
These relations have some important consequences:

\begin{itemize}
\item $C_{\mu}$ cannot be space-like, once $\mu ^2  =  - \tau ^2  \ge 0$;

\item $C_{\mu}=0$ yields $\mu=\tau=0$, so that no condensates would survive;

\item if $\mu=\tau=0$, then $C_{\mu}$ is light-like;

\item $C_{\mu}$ time-like implies $\mu\neq0$ and $\tau\neq0$. In this case, all condensates simultaneously contribute.
\end{itemize}

From (\ref{susy195}), we are ready to write down the photino dispersion relations,
\begin{equation}
\det \left( {J - LN^{ - 1} M} \right) = 0, \label{susy240}
\end{equation}
and the corresponding photon dispersion relations,
\begin{equation}
\det \left( {N - MJ^{ - 1} L} \right) = 0. \label{susy245}
\end{equation}

The fermionic opertor $J$ in (\ref{susy200}) is invertible. We compute $J^{-1}$ and quote its expression as follows:
\begin{equation}
J^{ - 1}  = A1_{4 \times 4}  + B\gamma _5  + R_\mu  \gamma ^\mu   + S_\mu  \gamma ^\mu  \gamma _5  + L_{\mu \nu } \Sigma ^{\mu \nu }, \label{susy250}
\end{equation}
with the coefficients $A, B, R_{\mu}, S_{\mu}$ and $L_{\mu \nu }=-L_{\nu \mu }$ listed below:

\begin{equation}
A=\left( {2{\mathop{\rm Re}\nolimits}\left( F\right) +\frac{\mu }{2}}\right)
\left( {4\left\vert F\right\vert ^{2}+\frac{3}{2}\mu ^{2}+2\delta -p^{2}+
\frac{{v^{2}}}{4}+\frac{1}{2}\left( {v\cdot C}\right) }\right) /\Delta ,
\label{susy255}
\end{equation}
\begin{equation}
B=i\left( {2{\mathop{\rm Im}\nolimits}\left( F\right) +i\frac{\tau }{2}}
\right) \left( {4\left\vert F\right\vert ^{2}+\frac{3}{2}\mu ^{2}+2\delta
-p^{2}+\frac{{v^{2}}}{4}+\frac{1}{2}\left( {v\cdot C}\right) }\right)
/\Delta ,  \label{susy260}
\end{equation}
\begin{equation}
R_{\mu }=\left[ {\left( {\frac{{p^{2}}}{2}+\frac{{v^{2}}}{8}+\frac{{\mu ^{2}}
}{4}+\frac{{\left( {v\cdot C}\right) }}{4}-2\left\vert F\right\vert
^{2}-\delta }\right) 2p_{\mu }-\frac{{\left\{ {\left( {p\cdot v}\right)
+\left( {p\cdot C}\right) }\right\} }}{2}\left( {v_{\mu }+C_{\mu }}\right) }
\right] /\Delta ,  \label{susy265}
\end{equation}
\begin{equation}
S_{\mu }=\left[ {\left( {\frac{{p^{2}}}{2}+\frac{{v^{2}}}{8}+\frac{3{\mu ^{2}}
}{4}+\frac{{\left( {v\cdot C}\right) }}{4}+2\left\vert F\right\vert
^{2}+\delta }\right) \left( {v_{\mu }+C_{\mu }}\right) -\left\{ {\left( {
p\cdot v}\right) +\left( {p\cdot C}\right) }\right\} p_{\mu }}\right]
/\Delta ,  \label{susy270}
\end{equation}
\begin{equation}
L_{\mu \nu }=\left[ {-2\left( {2{\mathop{\rm Im}\nolimits}\left( F\right) +
\frac{i}{2}\tau }\right) \left( {p_{\mu }v_{\nu }+p_{\mu }C_{\nu }}\right)
+\left( {2Re\left( F\right) +\frac{1}{2}\mu }\right) \left( {p_{\alpha
}v_{\beta }+p_{\alpha }C_{\beta }}\right) 
\varepsilon _{\mu \nu } \,^{\alpha \beta }}\right] /\Delta ,  \label{susy275}
\end{equation}
\begin{eqnarray}
\Delta &=& p^{4}-p^{2}\left[ {8\left\vert F\right\vert ^{2}+4\delta -\frac{{
v^{2}}}{2}-\left( {v\cdot C}\right) }\right] -2\left( {p\cdot v}\right)
\left( {p\cdot C}\right) -\left( {p\cdot v}\right) ^{2} \nonumber\\
&+&\left[ {4\left\vert
F\right\vert ^{2}+\frac{3}{2}\mu ^{2}+2\delta +\frac{{v^{2}}}{4}+\frac{{
\left( {v\cdot C}\right) }}{2}}\right] ^{2},  \label{susy280}
\end{eqnarray}
\begin{equation}
where \, \delta \equiv{\mathop{\rm Re}\nolimits}\left( F\right) \mu +i{\mathop{\rm Im}
\nolimits}\left( F\right) \tau.  \label{susy285}
\end{equation}

Now, that we know $J^{-1}$, we can rewrite the photino dispersion relation (\ref{susy240}) according to
\begin{equation}
\det \left( {J - LN^{ - 1} M} \right) = \left( {\det J} \right)\left[ {\det \left( {1 - J^{ - 1} LN^{ - 1} M} \right)} \right] = 0. \label{susy290}
\end{equation}
Since $(1 - J^{ - 1} LN^{ - 1} M)$ is invertible, the photino dispersion relation reduces to
\begin{equation}
\det J = \Delta  = 0, \label{susy295}
\end{equation}
with $\Delta$ given by (\ref{susy280}).
This expression then brings to light how the background vector and scalar, $v^{\mu}$ and $F$, and the fermion condensates, $\mu$, $\tau$ and $C^{\mu}$, combine to govern the photino propagation modes.

So long as the photon is concerned, its dispersion relation (\ref{susy245}) can be re-organized as ($N$ is invertible):
\begin{equation}
\det \left( {N - MJ^{ - 1} L} \right) = \left( {\det N} \right)\left[ {\det \left( {1 - N^{ - 1} MJ^{ - 1} L} \right)} \right] = 0. \label{disp5}
\end{equation}
Again, $({1 - N^{ - 1} MJ^{ - 1} L})$ is non-singular, so that
\begin{equation}
det N=0 \label{susy305}
\end{equation}
responds for the photon dispersion relation \cite{CFJ}:
\begin{equation}
p^4  + v^2 p^2  - \left( {v \cdot p} \right)^2  = 0, \label{susy310}
\end{equation}
from which the following masses come out: $m_{\gamma}=0$ and  $m_{\gamma}= |\vec v|$.

So, only the background vector $v^{\mu}$ actually affects the photon propagating modes. The scalar background, $F$, and the fermion condensates $\mu$, $\tau$, $C^{\mu}$, do not change the photon propagating modes of the non-supersymmetric Carroll-Field-Jackiw model. However, it is worthy mentioning that, even if the mixing operators, ${L}$ and $M$, in ${\cal O}$ do not contribute to both the photon and photino dispersion relations, we point out that they do affect the propagators of the photon-photino sector and they are very relevant for the analysis of the residue matrices of the $ 
\left\langle {\bar \Lambda _\alpha  \Lambda _\beta  } \right\rangle$-, $ 
\left\langle {\bar \Lambda _\alpha  A_\mu  } \right\rangle$- and $ 
\left\langle {A_\mu  A_\nu  } \right\rangle$-propagators at their poles. The latter are clearly the zeroes of the equations that give the dispersion relations, and this becomes clear since the propagators above can be read off from the matrix ${\cal O}^{-1}$ ( ${\cal O}$ given in eq.(\ref{susy195})), whose general form can be organized as follows:
\begin{equation}
{\cal O}^{ - 1}  = \left( {\begin{array}{*{20}c}
   X & Y  \\
   Z & W  \\
\end{array}} \right), \label{susy315}
\end{equation}
where
\begin{equation}
X \equiv \left( {J - LN^{ - 1} M} \right)^{ - 1}  = \left( {1 - J^{ - 1} LN^{ - 1} M} \right)^{ - 1} J^{ - 1}, \label{susy320}
\end{equation}
\begin{equation}
W \equiv \left( {N - MJ^{ - 1} L} \right)^{ - 1}  = \left( {1 - N^{ - 1} MJ^{ - 1} L} \right)^{ - 1} N^{ - 1} . \label{susy325}
\end{equation}
\begin{equation}
Y \equiv  - J^{ - 1} L\left( {N - MJ^{ - 1} L} \right)^{ - 1}  =  - J^{ - 1} L\left( {1 - N^{ - 1} MJ^{ - 1} L} \right)N^{ - 1}, \label{susy330}
\end{equation}
\begin{equation}
Z \equiv  - N^{ - 1} M\left( {J - LN^{ - 1} M} \right)^{ - 1}  =  - N^{ - 1} M\left( {1 - J^{ - 1} LN^{ - 1} M} \right)^{ - 1} J^{ - 1} . \label{susy335} 
\end{equation}
Eqs.(\ref{susy320}) and (\ref{susy325}) clearly confirm that the propagator poles, that are accomodated in $J^{-1}$ and $N^{-1}$, exactly correspond to the zeroes of dispersion relations (\ref{susy280}) and (\ref{susy310}). For the sake of our discussions in this work, we do not need to explicitly compute the propagators. We are only interested in working out the dispersion relations; this is why we do not carry out the explicit calculation of ${\cal O}^{-1}$.

To illustrate an astrophysical consequence of the emergence of a massive photon, as our dispersion relation shows,  we may compute the time delay between two electromagnetic waves of different frequencies, $\omega_{1}$ and $\omega_{2}$ \cite{de Broglie}. In SI units, for a source at a distance $l$ in equation ($16$) in \cite{Bonetti:2016vrq}, we get the time lapse as given below:
\begin{equation}
{\Delta _{CFJ}} = \frac{{l|{\bf V}{|^2}}}{{2c{\hbar ^2}}}\left( {\frac{1}{{w_1^2}} - \frac{1}{{w_2^2}}} \right)
. \label{susy335a} 
\end{equation}
As time delays are inversely proportional to the square of the frequency, we identify a massive photon in the spectrum. We have just seen that the mass comes out proportional to the breaking parameter $|{\bf v}|$. The comparison of the equation above with the corresponding expression for the de Broglie-Proca photon yields the identity equation ($18$) in \cite{Bonetti:2016vrq}. Finally, given the prominence of the delays of massive photon dispersion at low frequencies, a swarm of nano-satellites operating in the sub-MHz region \cite{Bentum} appears a promising avenue for improving upper limits through the analysis of plasma dispersion.

From (\ref{susy280}) and (\ref{susy310}), we see that only $v^{\mu}$ enters the photon dispersion relation though it also enters the photino dispersion relation. So, let us consider the particular situation
\begin{equation}
\Psi  = 0, \label{susy340}
\end{equation} 
and
\begin{equation}
F=0, \label{susy345}
\end{equation}
so that all fermion condensates are switched off. In such a case, the photino dispersion relation simplifies to
\begin{equation}
\Delta  = p^4  + \frac{1}{2}v^2 p^2  - \left( {v \cdot p} \right)^2  + \frac{1}{16}v^4. \label{susy350}
\end{equation}

It then becomes clear that a massless photon (according to (\ref{susy310}), characterized by $ {v \cdot p}=0$) is not  accompanied by a massless photino, since $ {v \cdot p}=0$ is not a zero of $\Delta$ whenever $p^{2}=0$. This confirms that the LSV actually induces a SUSY breaking, by splitting the photon and photino masses. In the special case of a space-like $v^\mu$, a massless photon is accompanied by a massive photino whose mass is calculated to be
\begin{equation}
{m_{\tilde \gamma }}  = \frac{1}{{ \sqrt{2} }} \left| {\vec v} \right|, \label{susy355}
\end{equation}
where $\vec v$ is the spatial component of $v^\mu$. In this particular situation, the photon-photino mass splitting is directly measured by $v^\mu$.

On the other hand, if $v^{\mu}=0$ and the fermions condensates are non-trivial, $p^{2}=0$ is always a zero of (\ref{susy310}) (so, a massless photon is present in the spectrum in such a case), but it is never a zero of $\Delta$, so that, in this special case, a massless photino never shows up, which is again compatible with the situation of broken SUSY:
\begin{equation}
{m_{\tilde \gamma }} = \bar \Psi \,\Psi. \label{susy355b}
\end{equation}

This result on the photino mass can be work out by computing the dispersion relation that follows from the field equations below:
\begin{equation}
\vec \nabla  \cdot \vec E - \vec v \cdot \vec B = i\sqrt 2 \, \vec \nabla  \cdot \left( {\bar \Psi {\gamma ^0}\vec \gamma {\gamma _5}\Lambda } \right), \label{susy355c}
\end{equation}

\begin{equation}
\vec \nabla  \times \vec E =  - {\partial _t}\vec B, \label{susy355d}
\end{equation}

\begin{equation}
\vec \nabla  \cdot \vec B = 0, \label{susy355e}
\end{equation}

\begin{equation}
\vec \nabla  \times \vec B - {v^0}\vec B + \vec v \times \vec E = {\partial _t}\vec E - i\sqrt 2\, \bar \Psi {\gamma ^0}\vec \gamma {\gamma _5}\,{\partial _t}\Lambda  + 2\sqrt 2 \,\bar \Psi \,{\Sigma ^{ij}}{\gamma _5}\, {\partial _j}\Lambda, \label{susy355f}
\end{equation}

\begin{equation}
i{\gamma ^\mu }{\partial _\mu }\Lambda  - 2{M_1}\Lambda  + 2i{M_2}{\gamma _5}\Lambda  + \frac{1}{2}{R_\mu }{\gamma ^\mu }{\gamma _5}\Lambda  =  - i\sqrt 2 {\gamma ^0}{\gamma _5}\vec \gamma \,\Psi  \cdot \vec E - \sqrt 2 {\gamma ^0}\vec \gamma \, \Psi  \cdot \vec B, \label{susy355g}
\end{equation}
where $\vec \gamma$ stands ${\gamma ^i}$ (recalling that ${\gamma _i} =  - {\gamma ^i}$), and to get the last line we used, ${\varepsilon _{ijk}}{\Sigma ^{ij}}{\gamma _5} =  - {\gamma ^0}{\gamma _k} = {\gamma ^0}{\gamma ^k}$, in the term $\left( {{\Sigma ^{\mu \nu }}{\gamma _5}\Psi {F_{\mu \nu }}} \right)$. We thus find that the photino mass is shifted by the fermion condensate and we get equation (\ref{susy355b}) ($\vec v=0$).

Having now clarified how both the photon and photino acquire that respective masses in terms of the components of the background superfield S in eq. (\ref{susy165}), it is a suitable moment to point out another connection between our model and an issue of potential astrophysical interest, namely, a sort of fermionic version of the Primakoff effect, in which a photon-photino conversion replaces the usual photon-axion conversion that characterizes the mentioned effect. Back to the component-field action of eq. (\ref{susy175}), let us consider its very last term, where the photino field and the photon field-strength are mixed through the background fermion, $\Psi$. This term may be viewed as a three-vertex that describes a photon-photino conversion promoted by $\Psi$. This means that the way we connect SUSY with LSV, as described by eqs. (\ref{susy135}) and (\ref{susy175}), opens up a possible path to describe the conversion of massive photons into massive photini. The latter are viable candidates to constitute part of the dark matter. Clearly, ours is not a realistic model; we are only inspecting the purely photonic sector, disconnected from the full-fledged SSM. However, we are highlighting the possibility that a supersymmetric Lorentz-symmetry-violating background, as described by the superfield action (\ref{susy135}), may accommodate an effect that describes how photons may be converted into a dark matter constituent.

Finally, we can work to get a photonic effective action by integrating out the photino field. To do that, we are allowed to redefine $\Lambda$ according to the shift:
\begin{equation}
\Upsilon=\Lambda +{{J}}^{-1}\sqrt{2}\Sigma^{\mu\nu}\gamma_5 \Psi F_{\mu\nu},  \label{susy360}
\end{equation}
and 
\begin{equation}
\bar{\Upsilon}=\bar{\Lambda} -\sqrt{2}\bar{\Psi}\gamma_5\Sigma^{\mu\nu}\bar{{{J}}}^{-1}F_{\mu\nu},  \label{susy365}
\end{equation}
where $J^{-1}$  and ${\bar J^{-1}}$ already explicitly computed. Though there is a manifest non-locality in the field reshufflings (\ref{susy360}) and (\ref{susy365}),
this is harmless so long as we are interested in reading off an effective action for the photon by eliminating the $\Lambda _\alpha$-$A^{\mu}$ mixing and integrating out the fermions in the action (\ref{susy175}).

With the explicit expressions for $J^{-1}$ and ${\bar J^{-1}}$ and by means of manipulations with the $\gamma ^{\mu }$-algebra, we are able to cast the form of the photon effective action as given below: 
\begin{eqnarray}
{\cal L} &=&  - \frac{1}{4}F_{\mu \nu }^2  + \frac{1}{4}\varepsilon ^{\mu \nu \alpha \beta } v_\mu  A_\nu  F_{\alpha \beta }  + F_{\mu \nu } \left( {\frac{1}{4}\mu A + \frac{1}{4}\tau B + \frac{1}{2}C_\rho  S^\rho  } \right)F^{\mu \nu }  + F_{\mu \lambda } \left( {2C^\mu  S_\nu} \right)F^{\lambda \nu }  \nonumber\\
&+& F_{\mu \nu } \left( {\frac{i}{2}\tau A + \frac{i}{2}\mu B - \frac{1}{2}C_\rho  R^\rho   - \frac{i}{8}\tau L_\rho ^\rho  } \right)\tilde F^{\mu \nu }.  \label{susy370}
\end{eqnarray}

It is remarkable to point out that the breaking of Lorentz
symmetry naturally induces axionic-like terms, $F\tilde{F}$, whose coefficients are originated from background fermion condensates. In (\ref{susy370}), we warn that, in all the coefficients, $A$, $B$, $R_{\mu}$, $S_{\mu}$ and $L_{\mu\nu}$, the terms where there appears a $4$-momentum, $p^{\mu}$ and $p^2$ are to be understood as written down in coordinate space ($p_{\mu}=i\partial _\mu$).

Next, we shall study the consequences of these condensates over the confining and screening phases of the above photon effective Lagrangian density. As already mentioned, we shall calculate the static potential using the gauge-invariant but path-dependent variables formalism. This can be done by computing the expectation value of the energy operator $H$ in the physical state $|\Phi\rangle$ describing the sources (${\langle H \rangle}_{\Phi}$) along the lines of Subsection $2.1$. 

The starting point is the foregoing effective Lagrangian density for $A_\mu$ for the special case $v^\mu=0$, $F\neq0$, $\delta\neq0$, $C_i=0$ and $C_0\neq0$, that is,
\begin{equation}
\mathcal{L} = - \frac{1}{4}F_{\mu \nu }  \! \left[ {\frac{{\nabla ^4 - a^2
\nabla ^2 + b^2 }}{{\nabla ^4 - m_1 \nabla ^2 + m_2^2 }}} \right] \! F^{\mu \nu } + \frac{{C_0^2 }}{2} F_{0i} \!\left[ {\frac{{\nabla ^2 + m_2 }}{{\nabla ^4 + m_1
\nabla ^2 + m_2^2 }}} \right] \! F^{0i} + \frac{Q}{2}F_{\mu \nu } \! \left[ {\frac{{
 {\nabla ^2 - m_2 } }} {{\nabla ^4 + m_1 \nabla ^2 + m_2^2 }}} \right]
\!  \tilde F^{\mu \nu },  \label{susy375}
\end{equation}
Here, $a^2 \equiv m_1 -
\left( {2P + C_0^2 } \right)$, $b^2 \equiv m_2 \left( {m_2 - C_0^2 }
\right) + 2 m_2P$, $m_1 \equiv 8\left| F \right|^2 + 4\delta$, $m_2
\equiv 4\left| F \right|^2 + \frac{3}{2}\mu ^2 + 2\delta$, $P \equiv \mu
\left( {2{\mathop{\rm
Re}\nolimits} \left( F \right) + \frac{\mu }{2}} \right) + i\left( {2{
\mathop{\rm Im}\nolimits} \left( F \right) + i\frac{\tau }{2}} \right)\tau$
and $Q \equiv 2i{\mathop{\rm Re} \nolimits} \left( F \right)\tau - 2{
\mathop{\rm Im}\nolimits} \left( F \right)\mu$. Recalling again that we restrict ourselves to the static potential, a consequence of this is that one can drop out terms with time derivatives in the Lagrangian density (\ref{susy375}). 

Here again, the quantization can be done in a similar manner to that in the Subsection $2.1$. This leads us to the Hamiltonian
 \begin{eqnarray}
H &=& \int {d^3 x}\Biggl[c(x) \left( {\partial _i \Pi ^i} \right) -  \frac{1}{2}\Pi _i
\left( {\frac{{\nabla ^4 - m_1 \nabla ^2 + m_2^2 }}{{\nabla ^4 - \xi ^2 \nabla ^2 + \rho ^2 }}} \right)\Pi ^i -\frac{Q}{2}\Pi _i \left( {\frac{{\nabla ^2 - m_2 }}{{\nabla ^4 - \xi ^2 \nabla ^2 + \rho ^2 }}} \right)B^i + QB_i \Pi ^i  \notag \\
&+& Q^2 B_i \frac{{\left( {\nabla ^2 - m_2 } \right)}}{{\left( {\nabla ^4 -
m_1 \nabla ^2 + m_2^2 } \right)}}B^i + \frac{1}{4}F_{ij}
\left( {\frac{{\nabla ^4 - a^2 \nabla ^2 + b^2 }}{{\nabla ^4 - m_1 \nabla ^2 +
m_2 ^2 }}} \right)F^{ij}\Biggr]\ ,  \label{susy380}
\end{eqnarray}
where $c(x) = c_1 (x) - A_0 (x)$, $\xi ^2 \equiv a^2 +C_0^2$ and $\rho ^2 \equiv b^2 - m_2C_0^2$.

Whereas that $\left\langle H \right\rangle _\Phi$ reads 
\begin{eqnarray}
\left\langle H \right\rangle _\Phi &=& \left\langle \Phi \right|\int {d^3 x} 
\Biggl[\frac{1}{2}\Pi _i\left( {\frac{{\nabla ^4 - m_1 \nabla ^2 + m_2^2 }}{{\nabla ^4 - \xi ^2 \nabla ^2 + \rho ^2 }}} \right)\Pi ^i -\frac{Q}{2}\Pi _i \left( {\frac{{\nabla ^2 - m_2 }}{{\nabla ^4 - \xi ^2 \nabla ^2 + \rho ^2 }}} \right)B^i + QB_i \Pi ^i  \notag \\
&+& Q^2 B_i \frac{{\left( {\nabla ^2 - m_2 } \right)}}{{\left( {\nabla ^4 -
m_1 \nabla ^2 + m_2^2 } \right)}}B^i + \frac{1}{4}F_{ij}
\left( {\frac{{\nabla ^4 - a^2 \nabla ^2 + b^2 }}{{\nabla ^4 - m_1 \nabla ^2 +
m_2 ^2 }}} \right)F^{ij}\Biggr]|\Phi\rangle .  \label{susy385}
\end{eqnarray}

By proceeding in the same way as in \cite{susy1}, we obtain the potential for two opposite charges
located at $\mathbf{y}$ and $\mathbf{y^{\prime }}$:
\begin{eqnarray}
V &=& - \frac{{q^2 }}{{4\pi \sqrt {1 - {\raise0.5ex\hbox{$\scriptstyle
{4\rho ^2 }$} \kern-0.1em/\kern-0.15em \lower0.25ex
\hbox{$\scriptstyle {\xi ^4
}$}}} }}\left\{ {\frac{{\left( {1 + \sqrt {1-{\raise0.5ex\hbox{$\scriptstyle
{4\rho ^2 }$} \kern-0.1em/\kern-0.15em \lower0.25ex
\hbox{$\scriptstyle {\xi ^4
}$}}} } \right)}}{2}\frac{{e^{ - M_1 L} }}{L} - \frac{{\left( {1 - \sqrt {1
- {\raise0.5ex\hbox{$\scriptstyle {4\rho ^2
}$}\kern-0.1em/\kern-0.15em \lower0.25ex\hbox{$\scriptstyle {\xi ^4
}$}}}} \right)}}{2}\frac{{e^{ - M_2 L} }}{L}} \right\}  \notag \\
&+& \frac{{q^2 }}{{8\pi \sqrt {\xi ^4 - 4\rho ^2 } }}\left\{ {\left( {m_1
M_1^2 - m_2^2 } \right)\ln \left( {1 + \frac{{\Lambda ^2 }}{{M_1^2 }}}
\right) + \left( {m_2 - m_1 M_2^2 } \right)\ln \left( {1 + \frac{{\Lambda ^2 
}}{{M_2^2 }}} \right)} \right\}L,  \label{susy390}
\end{eqnarray}
where $\Lambda$ is a cutoff and $|\mathbf{y}-\mathbf{y}^{\prime }|\equiv L$,
and $M_1^2 \equiv {\textstyle{\frac{1 }{2}}}\left( {\xi ^2 + \sqrt {\xi ^4
- 4\rho ^2 } } \right)$ and $M_2^2 \equiv {\textstyle{\frac{1 }{2}}}\left( {
\xi ^2 - \sqrt {\xi ^4 - 4\rho ^2 } } \right)$, $M_1\geq M_2$.

From this last expression it is clear that the effect of including
condensates is a linear potential, leading to the confinement of static
charges. Also, it may be noted that the same result is obtained in the
time-like case. 

Before concluding this Subsection, we discuss the meaning of the cutoff 
$\Lambda $. In this case, we first note that 
our effective model for the electromagnetic field is an effective
description that arises upon integration over the $\Lambda $-field, whose
excitations are massive (Recall that $\Gamma=0$ for $p^2=M_1^2$ and $
p^2=M_2^2$). Evidently, $1/M_1$ and $1/M_2$, are the Compton wavelengths
of these excitations, which define a correlation distance. We thus see that 
physics at distances of the order or lower than $1/M_2$ must necessarily take
 into account a microscopic description of axion fields. In other words, if we work with
energies of the order or higher than $M_2$, our effective description with
the integrated effects of $\Lambda $ is no longer sensible. Accordingly, 
we identify $\Lambda $ with $M_1$. Thus, finally we end up 
\begin{eqnarray}
V &=& - \frac{{q^2 }}{{4\pi \sqrt {1 - {\raise0.5ex\hbox{$\scriptstyle
{4\rho ^2 }$} \kern-0.1em/\kern-0.15em \lower0.25ex
\hbox{$\scriptstyle {\xi ^4
}$}}} }}\left\{ {\frac{{\left( {1 + \sqrt {1-{\raise0.5ex\hbox{$\scriptstyle
{4\rho ^2 }$} \kern-0.1em/\kern-0.15em \lower0.25ex
\hbox{$\scriptstyle {\xi ^4
}$}}} } \right)}}{2}\frac{{e^{ - M_1 L} }}{L} - \frac{{\left( {1 - \sqrt {1
- {\raise0.5ex\hbox{$\scriptstyle {4\rho ^2
}$}\kern-0.1em/\kern-0.15em \lower0.25ex\hbox{$\scriptstyle {\xi ^4
}$}}}} \right)}}{2}\frac{{e^{ - M_2 L} }}{L}} \right\}  \notag \\
&+& \frac{{q^2 }}{{8\pi \sqrt {\xi ^4 - 4\rho ^2 } }}\left\{ {\left( {m_1
M_1^2 - m_2^2 } \right)\ln \left( 2 \right) + \left( {m_2 - m_1 M_2^2 }
\right)\ln \left( {1 + \frac{{M_1^2 }}{{M_2^2 }}} \right)} \right\}L.
\label{susy395}
\end{eqnarray}

\subsection{Supersymmetric extension of the Carroll-Field-Jackiw model for electrodynamics II}

Our aim in this Subsection is to examine the case when LSV takes place in an environment dominated by SUSY. To this end, we shall focus on the case that the Lorentz symmetry is violated in the photon sector by an CPT-even $k_{F}$ term.

We start off with the action for the CPT-even term for the abelian gauge sector of Standard Model Extension:
\begin{equation}
S_{\mbox{CPT-even}}=-\frac{1}{4}\int d^4x
(k_F)_{\mu\nu\alpha\beta}F^{\mu\nu}F^{\alpha\beta}. \label{susy340}
\end{equation}

The tensor $k_F$, from now on written as $K_{\mu\nu\alpha\beta}$ displays the
properties: 
\begin{equation}
K_{\mu\nu\alpha\beta}=-K_{\nu\mu\alpha\beta}=-K_{\mu\nu\beta\alpha}=K_{\alpha
\beta\mu\nu}, \label{susy345}
\end{equation}
it is double-traceless and its fully anti-symmetric component is ruled out
because it yields a total derivative. As well-known, it depends on 19
parameters.
If moreover we wish to suppress the components that yield birefringence, we
end up with only 9 independent components. We shall consider here a particular
situation of the non-birefringent case, namely, the case in which we are left  with only  four  coefficients that signals violation of Lorentz symmetry; these are described by a four-vector, ($\xi _{\alpha }$). According to the ansatz discussed in \cite{Bailey:2004na,Betschart:2008yi}, we may finally parametrize $K_{\mu \nu \alpha\beta }$ as it follows below: 
\begin{equation}
K_{\mu \nu \alpha \beta }=\frac{1}{2}(\eta _{\mu \alpha }{\tilde{\kappa}}
_{\nu \beta }-\eta _{\mu \beta }{\tilde{\kappa}}_{\nu \alpha }+\eta _{\nu
\beta }{\tilde{\kappa}}_{\mu \alpha }-\eta _{\nu \alpha }{\tilde{\kappa}}
_{\mu \beta }), \label{susy350}
\end{equation}
\begin{equation}
{\tilde{\kappa}}_{\alpha \beta }=(\xi _{\alpha }\xi _{\beta }-\eta _{\alpha
\beta }\frac{(\xi _{\rho }\xi ^{\rho })}{4}), \label{susy355}
\end{equation}
and the essence of LSV is traced back to the constant background 4-vector $
\xi _{\mu }$, so that the $k_{F}$ action becomes 
\begin{equation}
S=\int d^{4}x\dfrac{1}{4}\left( \frac{1}{2}\xi _{\mu }\xi _{\nu
}F_{\,\,\,\kappa }^{\mu }F^{\kappa \nu }+\frac{1}{8}\xi _{\rho }\xi ^{\rho
}F_{\mu \nu }F^{{}}\mu \nu \right) .  \label{susy360}
\end{equation}

In our proposal, this is a more reasonable situation. If we were to identify
the whole tensor $K_{\mu\nu\alpha\beta}$ as a component of a given
superfield, higher spins (actually, $s= \frac{3}{2}$) would be present in a
global SUSY framework. Since we have $\xi_\mu$ as the signal of LSV, no risk
of higher fermionic spins in the background is undertaken if the effects of
the $K$-tensor are transferred to the $\xi^\mu$-vector.

In this paper, we shall be working with supersymmetry formulated in superspace and  in terms of  superfields. For that, we refer the reader to notations and conventions adopted in the work of ref. \cite{susy2}.

In the work of ref. \cite{susy2}, two ways have been suggested to
implement a SUSY-extension for a 4-vector background: $\xi_\mu$ may appear
as the gradient of a scalar (in this case, LSV is in a chiral superfield) or
a complete vector (with transverse and longitudinal components); in the
latter case, $\xi_\mu$ should be a vector component of what we call a vector
superfield. To consider a simpler fermionic set of partners, we choose to place 
$\xi^\mu$ in the chiral superfield: in the first case the supersymmetry is
implemented through a chiral multiplet and in the other by means of a vector
multiplet. For simplicity, we work only on the chiral case . In this proposal,
the extended action written in superfield formalism is: 
\begin{equation}
S^{(susy)}_{\mbox{CPT-even}}=\int d^4x d^2 \theta d^2 \bar{\theta}\,\, 
\Big{[}(D^\alpha \Omega) W_{\alpha} (\bar{D}_{\dot{\alpha}} \bar{ \Omega})
\bar{ W}^{\dot{\alpha}}+h.c\,\Big{]} =S_{ferm}+S_{boson}+S_{mixing},
\label{susy365}
\end{equation}
where the supersymetry covariant derivates, the superspace action and the superfields can be found in the work  \cite{susy2}, 
\begin{eqnarray}
{W_a}\left( x \right) &=& {\lambda _a}\left( x \right) + i\theta {\sigma ^\mu }\bar \theta {\partial _\mu }{\lambda _a}\left( x \right) + 2{\theta _a}D\left( x \right) - i{\theta ^2}{\left( {\bar \theta {\sigma ^\mu }} \right)_\alpha }{\partial _\mu }D\left( x \right) + \left( {{\sigma ^{\mu \nu }}\theta } \right){F_{\mu \nu }}\left( x \right) \nonumber\\
&-& \frac{1}{2}{\theta ^2}{\left( {{\sigma ^{\mu \nu }}{\sigma ^\rho }} \right)_\alpha }{\partial _\rho }{F_{\mu \nu }}\left( x \right) - i{\left( {{\sigma ^\mu }{\partial _\mu }\lambda \left( x \right)} \right)_\alpha }{\theta ^2}. \label{susy370}
\end{eqnarray}
is the well-known field-strength superfield ($\lambda$ is the photino, $
F_{\mu\nu}$ the usual gauge-field strength and $D$ the auxiliary field); the
chiral background superfield, $\Omega$, is $\theta$-expanded as follows 
\begin{equation}
\Omega \left( x \right) = S\left( x \right) + \sqrt 2 \theta \zeta \left( x \right) + i\theta {\sigma ^\mu }\bar \theta {\partial _\mu }S\left( x \right) + {\theta ^2}G\left( x \right) + \frac{i}{{\sqrt 2 }}{\theta ^2}\bar \theta {\bar \sigma ^\mu }{\partial _\mu }\zeta \left( x \right) - \frac{1}{4}{\bar \theta ^2}{\theta ^2}S\left( x \right), \label{susy375}
\end{equation}
where $S$ and $G$ are complex scalars and $\zeta$ is a Weyl component of a
Majorana fermion. By projecting the action (6) into component fields, we
readily get that $\xi_\mu=\partial_\mu S$ and the $S_{boson}$, $S_{ferm}$
and $S_{mixing}$ may be found, in terms of Weyl spinors in ref. \cite{susy2}. We
prefer to quote below the component-field action directly in terms of
Majorana spinors, for it is much simpler and one can control much more
easily the various couplings present in the action.

At this point, we also make a special consideration about the background
superfield $\Omega$: taking $S$ linear in $x^\mu$ ($S=\xi_\mu x^\mu$, $
\xi_\mu$ constant), $\partial_\mu \zeta=0$ and $G=0$, is compatible with
SUSY, in the sense that these properties are kept if global SUSY
transformations are done, and moreover we reproduce the $k_F$-term as we
wish from the very beginning. Now, we shall move on with two purposes:

\begin{itemize}
\item (i) to rewrite the whole action in terms of 4-components Majorana
spinors, $Z\equiv(\zeta \,\,\,\bar{\zeta})^t$ and $\Lambda
\equiv(\lambda\,\,\, \bar{\lambda})^t$,

\item (ii) to Fierz-Rearrange the terms in $S_{ferm}$ where the fermions $
\zeta$ and $\lambda$ are mixed. This process selects for us 3 types of
background fermion condensates (already written in terms of Majorana
spinors): 
\begin{eqnarray}
\Theta=&&\bar{Z}Z  \notag \\
\tau=&&\bar{Z}\gamma_5 Z  \notag \\
C^{\mu}=&&\bar{Z}\gamma^\mu \gamma_5 Z,  \label{susy380}
\end{eqnarray}
\end{itemize}

for which the relations below hold true: 
\begin{equation}
\Theta^2=-\tau^2=\frac{1}{4}C^\mu C_\mu ,  \notag
\end{equation}
\begin{equation}
\Theta \tau=0\,\,\,\,\mbox{and}\,\,\,\,\Theta C^\mu= \tau C^\mu=0. \label{susy385}
\end{equation}

With all these considerations, the action (\ref{susy365}) can be brought into
a more readable form: 
\begin{eqnarray}
S_{boson}&=& \int d^4x \Big[ D^2 (32|G|^2+16\partial_\mu S \partial^\mu
S^*)+8iD F^{\mu\nu}(\partial_\mu S\partial_\nu S^*- \partial_\mu
S^*\partial_\nu S)  \nonumber\\
&-&8F^{\mu\kappa}F_{\kappa}^{\,\,\,\,\nu}(\partial_\mu S\partial_\nu S^*+
\partial_\mu S^*\partial_\nu S)-4F^{\mu\nu}F_{\mu\nu} \partial_\alpha S
\partial^\alpha S^*\,\Big];  \label{susy390} 
\end{eqnarray}

\begin{equation}
{S_{ferm}} = \int {{d^4}x} \left( {{C^\mu }\bar \Lambda {\gamma ^\nu }{\gamma _5}{\partial _\mu }{\partial _\nu }\Lambda  + y{C_\mu }\bar \Lambda {\gamma ^\mu }{\gamma _5}\Lambda } \right), \label{susy395} 
\end{equation}
where $y = \frac{{4 - \sqrt 2 }}{{16}}$.

\begin{eqnarray}
{S_{mixing}} \!\!\!\!&=& \!\!\!\!\int {{d^4}x} D\left( {10\sqrt 2 {\mathop{\rm Re}\nolimits} \left( {{\partial _\mu }S} \right)\left( {\bar Z{\partial _\mu }\Lambda } \right) - 8\sqrt 2 i{\mathop{\rm Re}\nolimits} \left( {{\partial _\mu }S} \right)\left( {\bar Z{\Sigma ^{\mu \nu }}{\partial _\nu }\Lambda } \right)8\sqrt 2 {\mathop{\rm Im}\nolimits} \left( {{\partial _\mu }S} \right)\left( {\bar Z{\Sigma ^{\mu \nu }}{\gamma _5}{\partial _\nu }\Lambda } \right)} \right) \nonumber\\
 &+& \!\!\!\!\int {{d^4}x} D10\sqrt 2 i{\mathop{\rm Im}\nolimits} \left( {{\partial _\mu }S} \right)\left( {\left( {\bar Z{\gamma _5}{\partial ^\mu }\Lambda } \right)} \right) \nonumber\\
 &+&\!\!\!\! \int {{d^4}x} \left( { - 3\sqrt 2 {\mathop{\rm Im}\nolimits} \left( {{\partial _\nu }S} \right)\left[ {{\partial _\mu }{F^{\mu \nu }}} \right]\bar Z\Lambda  + 3\sqrt 2 {\mathop{\rm Re}\nolimits} \left( {{\partial _\nu }S} \right)\left[ {{\partial _\mu }{F^{\mu \nu }}} \right]\bar Z{\gamma _5}\Lambda } \right) \nonumber\\
 &+&\!\!\!\! \int {{d^4}x} \left( {4\sqrt 2 i{\partial _{[\nu }}{F_{\mu ]\alpha }}{\mathop{\rm Im}\nolimits} \left( {{\partial ^\alpha }S} \right)\bar Z{\Sigma ^{\mu \nu }}\Lambda  + 4\sqrt 2 i{\partial _{[\mu }}{{\tilde F}_{\nu ]\alpha }}{\mathop{\rm Re}\nolimits} \left( {{\partial ^\alpha }S} \right)\bar Z{\Sigma ^{\mu \nu }}\Lambda } \right) \nonumber\\
 &+&\!\!\!\! \int {{d^4}x} \left( {4\sqrt 2 {\partial _{[\nu }}{F_{\mu ]\alpha }}{\mathop{\rm Im}\nolimits} \left( {{\partial ^\alpha }S} \right)\bar Z{\Sigma ^{\mu \nu }}{\gamma _5}\Lambda  + 4\sqrt 2 {\partial _{[\nu }}{{\tilde F}_{\mu ]\alpha }}{\mathop{\rm Re}\nolimits} \left( {{\partial ^\alpha }S} \right)\bar Z{\Sigma ^{\mu \nu }}{\gamma _5}\Lambda } \right), \label{susy400}
\end{eqnarray}
where we adopt that the indices enclosed  by square brackets stand for anti-symmetrization, $\tilde{F}_{\mu\nu}$ is the dual of $F_{\mu\nu}$ and $\Sigma_{\mu\nu}=\frac{i}{4}[\gamma_\mu,\gamma_\nu]$.
$D$ appears as an auxiliary field and we are going to
eliminate it below upon use of its corresponding equation of motion.

The equations above are indeed more manageable to work with. In order to complete
our model, we must add  to equations (\ref{susy390})-(\ref{susy400}) the supersymmetric version of
the Maxwell action. After this is done, its advisable to  eliminate
the auxiliary field, $D$, by means of algebraic equation of motion. Notice
that the total action can be written in terms of auxiliary field in the form
below: 
\begin{equation}
S^{(full)}=S^{(susy)}_{\mbox{Maxwell}}+S^{(susy)}_{\mbox{CPT-even}}=S+\int
d^4x \,\beta D +\int d^4x \,\alpha D^2, \label{susy405}
\end{equation}
\begin{equation}
S^{(full)}=S-\int dx^4 \,\frac{\beta^2}{2(2+\alpha)}, \label{susy410}
\end{equation}
where $\alpha$ and $\beta$ are expressed in terms of background and fields
in the gauge sector as follows: 
\begin{eqnarray}
\alpha=&&16(\partial_\kappa S \partial^\kappa S^*),  \notag \\
\beta{}=&&10\sqrt{2} Re(\partial_\mu S)(\bar{Z}\partial_\mu \Lambda)-8\sqrt{2
}i Re(\partial_\mu S) (\bar{Z}\Sigma^{\mu\nu}\partial_\nu \Lambda) +  \notag
\\
&&+ 8\sqrt{2} Im(\partial_\mu S)(\bar{Z}\Sigma^{\mu\nu}\gamma_5 \partial_\nu
\Lambda)+10\sqrt{2}i Im(\partial_\mu S)(\bar{Z}\gamma_5 \partial^\mu
\Lambda)+16\,m_{\mu\nu}F^{\mu\nu},  \label{susy415}
\end{eqnarray}
where $m_{\mu\nu}=Re(\partial_\mu S)Im(\partial_\nu S)-Re(\partial_\nu
S)Im(\partial_\mu S)$.\newline

The calculation of $\beta^2$ involves again the use of Fierz identities and
properties of bilinear formed by anticommuting Majorana spinors. The final
result is somehow cumbersome, thus we refer the reader to Appendix A 
of ref. \cite{susy2}.

Then, incorporating the $\beta^2$-term into the action, we have: 
\begin{eqnarray}
{S^{\left( {full} \right)}} &=& \int {{d^4}x} \left( { - \frac{1}{4}{F_{\mu \nu }}{F^{\mu \nu }} - \frac{1}{4}{K_{\mu \nu \alpha \beta }}{F^{\mu \nu }}{F^{\alpha \beta }} - \frac{{64}}{{\left( {1 + 8{\partial _\rho }S{\partial ^\rho }{S^ * }} \right)}}{m_{\mu \nu }}{m_{\alpha \beta }}{F^{\mu \nu }}{F^{\alpha \beta }}} \right) \nonumber\\
&+&\int {{d^4}x}  - \left( {\bar \Lambda \frac{{\tilde a}}{{4\left( {1 + 8{\partial _\kappa }S{\partial ^\kappa }{S^ * }} \right)}}\Lambda  + \bar \Lambda \frac{{\tilde b}}{{4\left( {1 + 8{\partial _\kappa }S{\partial ^\kappa }{S^ * }} \right)}}{\gamma _5}\Lambda } \right) \nonumber\\
 &+& \int {{d^4}x}  - \left( {\bar \Lambda \left[ {\left( {C \cdot \partial } \right){\partial _\mu } + y{C_\mu } + {C^\alpha }{d_{\alpha \mu }}} \right]{\gamma ^\mu }{\gamma _5}\Lambda  + 2\bar ZN\Lambda } \right). \label{susy420}
\end{eqnarray}
The $K_{\mu\nu\alpha\beta}$-tensor appearing above, now in the supersymetric background,  is given in terms of the complex vectors  $\xi_\mu$ according to  the expression below:
\begin{equation}
 K_{\mu\nu\alpha\beta}=-16(\eta _{\mu \alpha }{\tilde{\kappa}}
_{\nu \beta }-\eta _{\mu \beta }{\tilde{\kappa}}_{\nu \alpha }+\eta _{\nu
\beta }{\tilde{\kappa}}_{\mu \alpha }-\eta _{\nu \alpha }{\tilde{\kappa}}
_{\mu \beta }), \label{susy425}
\end{equation}
with:
\begin{equation}
 {\tilde{\kappa}}_{\alpha \beta }=\dfrac{1}{2}(\xi _{\alpha }\xi^* _{\beta }+\xi^* _{\alpha }\xi _{\beta })-\frac{\eta _{\alpha
\beta }}{4}(\xi _{\rho }\xi ^{\rho \,*}), \label{susy430}
\end{equation}
also, we should point out that there is an extra contribution to the $FF$-term given by the coefficients $m_{\alpha\beta}m_{\mu\nu}$. This latter contribution is intrinsic to supersymetry. It is important here to remark  that, even though $\xi_\mu$ is complex, as imposed by supersymetry, the $K$-tensor is automatically real, as it should be to avoid dissipating solutions to the fields equations.\\

The coefficients  $d_{\alpha\mu}$, ${\tilde{a}}$ and ${
\tilde{b}}$ can all be found in \cite{susy2}. The $N$-matrix above, that mixes
the background fermion and the photino, is given by a lengthy expression
that involves the photon field and its field strength, $F_{\mu\nu}$. This term
mixes therefore the photon and the photino fields, and the explicit form of $
N$ follows below: 
\begin{subequations}
\begin{align}
N=&I^{(1)}+iI^{(2)}\gamma_5+iI_{\mu\nu}\Sigma^{\mu\nu}, \\
&\mbox{where}  \notag \\
I^{(1)}=&-\frac{3}{2}\sqrt{2} Im(\partial_\mu S) \partial_\alpha
F^{\alpha\mu}+\dfrac{20\sqrt{2}}{(1+8 \partial_\kappa S \partial^\kappa S^*)}
m_{\alpha\beta}Re(\partial_\rho S) \partial^\rho F^{\alpha\beta}, \\
I^{(2)}=&\frac{3}{2}\sqrt{2} Re(\partial_\mu S) \partial_\alpha
F^{\alpha\mu}+\dfrac{20\sqrt{2}i}{(1+8 \partial_\kappa S \partial^\kappa S^*)
}m_{\alpha\beta}Im(\partial_\rho S) \partial^\rho F^{\alpha\beta}, \\
I_{\mu\nu}=&2\sqrt{2}\Big{[} Im(\partial^\alpha S)
\partial_{[\nu}F_{\mu]\alpha}+Re(\partial^\alpha S) \partial_{[\nu}\tilde{F}
_{\mu]\alpha}\Big{]}-\dfrac{16\sqrt{2}}{(1+8 \partial_\kappa S
\partial^\kappa S^*)}m_{\alpha\beta}Re(\partial_\mu S) \partial_\nu
F^{\alpha\beta}  \notag \\
&\sqrt{2} \epsilon_{\alpha\beta\mu\nu}\Big{[} Re(\partial_{\rho} S
)\partial^\alpha \tilde{F}^{\beta\rho}-Im (\partial_{\rho} S )
\partial^\alpha\tilde{F}^{\beta\rho}\Big{]}-\dfrac{8\sqrt{2}}{(1+8
\partial_\kappa S \partial^\kappa S^*)}\epsilon_{\alpha\beta\mu\nu}m_{\kappa
\lambda}Re(\partial^\alpha S) \partial^\beta F^{\kappa\lambda}.
\end{align}

Let us call the reader's attention to the fact that the $A^\mu-\Lambda$ mixed
term appears in the form $\bar{Z} N \Lambda$; the $N$-matrix is written in
terms of $1$, $\gamma_5$ and $\Sigma_{\mu\nu}$, and the coefficients $I^{(1)}
$, $I^{(2)}$ and $I_{{\mu\nu}}$ contain terms in the background field $S$
(through $\partial_\mu S$) and $F_{\mu\nu}$. As a whole, the term $\bar{Z} N
\Lambda$ is quadratic in the bosonic background and quadratic (but
non-diagonal) in the degrees of freedom of the gauge sector ($A^\mu$ and $
\Lambda$).

The $N$-matrix previously defined depends on the field strength, $
F^{\alpha\beta}$, through terms of the form $\partial^\mu F^{\alpha\beta}$.
For convenience, we factor out $N$ according to  the following splitting:
$N=N^{^{\prime }}_{\alpha}A^{\alpha}$. $N^{'}$ is therefore a combination of differential operators acting on the gauge potential $A_\mu$ according to the expression for $N$. This allows us to rewrite in a more
compact form the quadratic action in the photon and photino fields. We unify
the latter in a sort of doublet: $\Psi\equiv (
\begin{array}{c}
\Lambda \\ 
A_\nu
\end{array}
)$, $\bar{\Psi}\equiv (\bar{\Lambda} \,\,\,A_\mu)$, so that the full action
may be brought into the form 
\end{subequations}
\begin{equation}
S^{\mbox{(full)}}=\frac{1}{2}\int dx^4 \bar{\Psi} {\mathcal{O}} \Psi, \label{susy435}      \\
\end{equation}
where the matrix operator $\mathcal{O}$ is given by 
\begin{equation}
{\mathcal{O}}= \left( 
\begin{array}{cc}
M & \,\,\,N^{^{\prime }} \\ 
N^{^{\prime }} & \,\,\,Q \\ 
\end{array}
\right),
\end{equation}
with the sub-matrices given as below: 
\begin{eqnarray}
M &=&  - \frac{{\tilde a}}{{4\left( {1 + 8{\partial _\kappa }S{\partial ^\kappa }{S^ * }} \right)}}{1_{4 \times 4}} - \frac{{\tilde b}}{{4\left( {1 + 8{\partial _\kappa }S{\partial ^\kappa }{S^ * }} \right)}}{\gamma _5}  \nonumber\\
 &+& i\frac{{{\gamma ^\mu }{\partial _\mu }}}{2} + \left( { - \left( {C \cdot \partial } \right){\partial _\mu } + y{C_\mu } + {C^\alpha }{d_{\alpha \mu }}} \right){\gamma ^\mu }{\gamma _5}, \label{susy440}
\end{eqnarray}
\begin{equation}
{Q_{\mu \nu }} =  - \frac{1}{2}{\theta _{\mu \nu }} + \left( {{J_{\mu \alpha \beta \nu }} - {J_{\mu \alpha \nu \beta }} + {J_{\alpha \mu \nu \beta }} - {J_{\alpha \mu \beta \nu }}} \right){w^{\alpha \beta }}, \label{susy445}
\end{equation}
where
\begin{equation}
{J_{\mu \alpha \beta \nu }} =  - \frac{1}{4}{K_{\mu \alpha \beta \nu }} - \frac{{64}}{{\left( {1 + 8{\partial _\rho }S{\partial ^\rho }{S^ * }} \right)}}{m_{\mu \alpha }}{m_{\beta \nu }}. \label{susy450}
\end{equation}
The quantities $\theta_{\mu\nu}$, $\omega_{\mu\nu}$ and $d_{\mu\nu}$ can be found in Appendix B of ref. \cite{susy2}.

A conventional
procedure would consist in explicitly calculating $\mathcal{O}^{-1}$ in
order to get the  $\bar{\Lambda} \Lambda-$, $\Lambda A_\mu-$ and $
A_\mu A_\nu-$ propagators, whose pole structure corresponds to the
dispersion relation. However, if we are simply interested in the dispersion
relations for the photon and photino fields, we can concentrate only on the
matrices $M$ and $Q$, as was  shown in more details in the paper of ref.
\cite{susy2}. Actually, the poles of the photon and photino propagators can
be read off from $\mbox{det} Q=0$ and $\mbox{det} N =0$, respectively.

The photino propagator corresponds to the inverse matrix $M^{-1}$, whose
pole structure is found in det $M$: 
\begin{equation}
M^{-1}=A+B\gamma_5+v_\mu \gamma^\mu+\omega_\mu
\gamma^\mu\gamma_5, \label{susy455}
\end{equation}
with the coefficients given by (in the momentum space): 
\begin{equation}
A=\dfrac{\tilde{a}p^2}{16(1+8 \partial_\kappa S \partial^\kappa S^*)\Delta}, \label{susy460}
\end{equation}
\begin{equation}
B=-\dfrac{\tilde{b}p^2}{16(1+8 \partial_\kappa S \partial^\kappa S^*)\Delta}, \label{susy465}
\end{equation}
\begin{equation}
v_\mu=\Big{[}\dfrac{\tilde{a}^2-\tilde{b}^2}{16(1+8 \partial_\kappa S
\partial^\kappa S^*)^2}-\frac{p^2}{4}-\tilde{w}^2 \Big{]}\frac{p_\mu}{2\Delta}+
\dfrac{(\tilde{w}.p)\tilde{w}_\mu}{\Delta}, \label{susy470}
\end{equation}
\begin{equation}
\omega_\mu=(1-y)p^2(p.C)\dfrac{p_\mu}{2\Delta}+(C^\alpha p^\beta
d_{\alpha\beta})\dfrac{p_\mu}{2\Delta}-\dfrac{p^2}{4\Delta}\Big{[}
(p.C)p_\mu-y p^2 C_\mu+C^\alpha d_{\alpha\mu}\Big{]}, \label{susy475}
\end{equation}
and 
\begin{equation}
\Delta=\frac{p^4}{16}-(p.\tilde{w})^2-\dfrac{p^2\tilde{a}^2}{32(1+8
\partial_\kappa S \partial^\kappa S^*)^2}+\dfrac{p^2\tilde{b}^2}{32(1+8
\partial_\kappa S \partial^\kappa S^*)^2}+\frac{p^2}{2}\tilde{w}^2, \label{susy480}
\end{equation}
where ${p}^{\mu}$ is the photino momentum and the coefficients $\tilde{a}$, $\tilde{b}$ and $\tilde{w}_\mu$ are explicitly stated in eqs. [58-61] of Appendix B of our work \cite{susy2}.

We can separate the denominator $\Delta$ in two parts: one containing terms
up to 2nd. order in powers of $\partial_\mu S$ and  another piece that
only contains higher powers in $\partial_\mu S$ . This splitting is suitable
if we recall that the LSV parameters are very tiny, so that we confine our
considerations to terms which are second order in $\partial_\mu S$, and we
collect higher terms in ${\mathcal{O}}(3)$: 
\begin{equation}
\Delta=p^4\Theta^2\tilde{\Delta}=p^4\Theta^2\Big{(}\frac{1}{16\Theta^2}+
\left[ C^{(1)}p^2+C^{(2)}_{\mu\nu}p^\mu p^\nu \right]+\mathcal{O}(3)\Big{)},
\label{susy485}
\end{equation}
where 
\begin{equation}
C^{(1)}=(y^2-y-\frac{1}{2})+\left[\dfrac{1}{(1+8\partial_\mu S \partial^\mu
S^*)} \right](4y-2)(\eta_{\mu\nu}t^{\mu\nu}),  \label{susy490}
\end{equation}
\begin{equation}
C^{(2)}_{\mu\nu}=\left[\dfrac{1}{2(1+8\partial_\mu S \partial^\mu S^*)} 
\right][42y-29]t_{\mu\nu}.  \label{susy495}
\end{equation}

Since $K_{\mu\nu\alpha\beta}$ is a linear combination of bilinear in $
\partial_\mu S$, terms of ${\mathcal{O}}(3)$ or higher in eq. (15) are
discarded . We also notice that the coefficient $C^{(2)}_{\mu\nu}$ is much
smaller than $C^{(1)}$ since $|t_{\mu\nu}|<< 1$, so, in this approximation,
it is possible to remove the term that mixes the momenta and we find a very
simple dispersion relation for the photino which is given by 
\begin{equation}
\Delta^{(approx)}= {\Theta ^2}{p^4}\left[ {\frac{1}{{16{\Theta ^2}}} + {C^{\left( 1 \right)}}{p^2}} \right],  \label{susy500}
\end{equation}
with 
\begin{equation}
{m_{\tilde \gamma }^{2}} = - \frac{1}{{16{\Theta ^2}}C^{\left( 1 \right)}},
\label{susy505}
\end{equation}
notice that $C^{(1)}$ is negative. Here, contrary to the Carrol-Field-Jackiw
supersymmetrized model of ref. \cite{Belich:2003fa}, the photino mass carries an
explicit dependence on the $\Theta$-fermion condensate. This is a new
feature of the $k_F$-model. We highlight here that even if the bosonic part of the background (the four-vector $\xi_\mu$) is trivial, the photino mass does not vanish because it is a natural consequence of the condensation of the fermionic sector of the background. This is a very salient aspect of the connection between supersimmetry and the violation of Lorentz covariance. 

Following along analogous steps, we are able to find the dispersion relation
for the photon 
\begin{equation}
p_{\pm }^{0}=(1+\rho \pm \sigma )|\bar{p}|,  \label{susy510}
\end{equation}
where $\rho =\frac{1}{2}\tilde{K}_{\alpha }^{\,\,\,\alpha }$ and $\sigma
^{2}=\frac{1}{2}(\tilde{K}_{\alpha \beta })^{2}-\rho ^{2}$, with $\tilde{K}
^{\alpha \beta }=K^{\alpha \beta \mu \nu }\hat{p}_{\mu }\hat{p}_{\nu }$ and $
\hat{p}^{\mu }=p^{\mu }/|\bar{p}|$ \cite{Bonetti:2017toa}.

Finally, by eliminating the mixed $A_\mu \, \Lambda$ terms, we shall find an
effective action for the purely photonic sector . In the action, the term that
combines these fields is given by $2 \bar{Z}N \Lambda$. We notice that this
term can be removed  by performing a convenient shift in the photino field.
By redefining the fermion field according to $\Upsilon=\Lambda+M^{-1}\bar{N}Z
$, we attain a new action that is totally diagonal in the fields $\Upsilon$
and $A_\mu$. With the help of the properties of the fermionic condensates
(7) and the gamma-matrix algebra, the redefinition of $\Lambda$ suggested
above yields an effective term for the photon sector which can be expressed
as follows: 
\begin{eqnarray}
S_{\mbox{effective}}^{\,(photon)}=&&\int d^4 x \bar{Z}(N M^{-1} \bar{N})Z 
\notag \\
=&&\int d^4 x\Big{[} (I^{(1)}I^{(1)}-I^{(2)}I^{(2)}+\frac{1}{2}
I_{\mu\nu}I^{\mu\nu})(A\Theta+B\tau)+i(2I^{(1)}I^{(2)}-\frac{1}{2}I_{\mu\nu}
\tilde{I}^{\mu\nu})(A\tau+B \Theta)+  \notag \\
&&(I^{(1)}I^{(1)}+I^{(2)}I^{(2)}+\frac{1}{2}I_{\mu\nu}I^{\mu\nu})\omega_
\rho C^\rho+2I^{(1)}I^{\kappa\rho}\omega_\kappa C_{\rho} -2I^{(2)}\tilde{
I}^{\kappa\rho}\omega_\kappa C_{\rho}\Big{]},  \label{susy515}
\end{eqnarray}
where the coefficients $I^{(1)}, I^{(2)}$ and $I_{\mu\nu}$ ( $\tilde{I}^{\mu\nu}$ is the dual tensor of $I_{\mu\nu}$) only exhibit
derivatives of the field strength.

Taking into account the previous discussion on the approximation we adopt to
treat the LSV parameters, we can also ignore the terms of order $\mathcal{O}
(3)$ in Eq.(16), so that the full effective Lagrangian density (in the
momentum space) for the photon is given by the expression 
\begin{equation}
\mathcal{L}=\mathcal{L}_{\mbox{old}}+\mathcal{L}_{\mbox{effective}},
\label{susy520}
\end{equation}
where 
\begin{equation}
\mathcal{L}_{\mbox{old}}=-\frac{1}{4}F_{\mu\nu}F^{\mu\nu}-16
t_{\mu\nu}F^{\mu\kappa}F_\kappa^{\,\,\,\nu}-4
F_{\mu\nu}F^{\mu\nu}(t_{\alpha\beta}\eta^{\alpha\beta}),  \label{susy525}
\end{equation}
and 
\begin{equation}
\mathcal{L}_{\mbox{effective}}=\dfrac{1}{\tilde{\Delta}}(\frac{y}{4}-\frac{1
}{8})t_{\rho\lambda}\big[4p^2
F_{\mu}^{\,\,\,\,\rho}F^{\mu\lambda}-\eta^{\rho\lambda} p^2
F_{\mu\nu}F^{\mu\nu} \big]+ \dfrac{1 }{\tilde{\Delta}}(\frac{5y}{8}+\frac{13
}{16})t_{\rho\lambda} \big[ p_\mu p_\nu F^{\mu\rho} F^{\nu\lambda}\big].
\label{susy530}
\end{equation}
It is to be highlighted here a remarkable difference between the effective photonic
actions derived in the Carroll-Field-Jackiw and in the $k_F$-cases in a SUSY scenario:
the supersymmetric version of the $k_{AF}$-term induces a purely photonic action with
CP-violating axionic terms; the $k_F$-case, on the other hand, does not induce CP-breaking
terms in the effective photonic action. As we can check in the eq. (\ref{susy500}) above, no CP-
violating term of the form $F\tilde F$ shows up.  It is also to be noticed also that in both, the 
$k_{AF}$-  and  the $k_F$-cases, $\partial F$-terms appear, so that, the $k_F$-  and  the $k_{AF}$-models are equally sensitive in high-frequency regime. The most remarkable difference is actually the
absence of CP-violating terms in the photonic action of the $k_F$-case. Let us mention here that  ${t_{\mu \nu }}$  is defined in Appendix B of ref. \cite{susy2}.

Following our earlier procedure, we now turn our attention to the calculation of the interaction energy between static point-like sources for the effective model under consideration.
In such a case the effective Lagrangian density reads
\begin{eqnarray}
\mathcal{L} &=& - \frac{1}{4}F_{\mu \nu } \left[ {1 + 16t_\alpha ^\alpha -
4\left( {\frac{y}{4} - \frac{1}{8}} \right)t_\alpha ^\alpha \frac{\Delta }{{
\tilde \Delta }}} \right]F^{\mu \nu } -16t_{\mu \nu } F^{\mu \lambda }
F^{\nu}\ _{\lambda} - 4\left( {\frac{y}{4} - \frac{1}{8}} \right)t_{\rho
\lambda } F^{\mu \lambda } \frac{\Delta }{{\tilde \Delta }}F_\mu ^\rho 
\notag \\
&-& \left( {\frac{{5y}}{8} + \frac{{13}}{{16}}} \right)t_{\rho \lambda }
F^{\mu \lambda } \frac{{\partial _\mu \partial _\nu }}{{\tilde \Delta }}
F^{\nu \rho },  \label{susy535}
\end{eqnarray}
where $\Delta \equiv \partial _\mu \partial ^\mu$. Once again, in this 
static case, we shall replace $\Delta$ by $-\nabla ^2
$ in Eq.(\ref{susy505}). In passing we recall that the only
non-vanishing $t_{\mu \nu }$-terms are the diagonal ones, since 
$t_{\mu \nu }$ can be brought into a diagonal form. In the same way,
without loss of generality, we may always choose $t_{00}\ne 0$.

Then, the effective Lagrangian takes the form 
\begin{equation}
\mathcal{L} = - \frac{1}{4}\gamma F_{\mu \nu } \frac{{\left( {\nabla ^2 -
M^2 } \right)}}{{\left( {\nabla ^2 - m^2 } \right)}}F^{\mu \nu } +16t_{0 0 }
F_{i 0 } F^{i0} - \frac{{A_4 }}{{A_2 }}t_{00} F_{i 0 } \frac{{\nabla ^2 }}{{
\left( {\nabla ^2 - m^2 } \right)}}F^{i 0} - \frac{{A_5 }}{{A_2 }}t_{00}
F^{i 0} \frac{{\partial _i \partial _j }}{{\left( {\nabla ^2 - m^2 } \right)}
}F^{j 0}.  \label{susy540}
\end{equation}
Here we have simplified our notation by setting $\gamma = \frac{{A_3 A_2 - A_4 t_\alpha ^\alpha }}{{A_2 }}$, $M^2 = 
\frac{{A_3 A_1 }}{{A_3 A_2 - A_4 t_\alpha ^\alpha }}$, $m^2 = \frac{{A_1 }}{{
A_2 }}$, $A_1 = \frac{1}{{16\Theta ^2 }}$, $A_2 = -
C^{\left( 1 \right)}$, $A_3 = \left( {1 + 16t_\alpha ^\alpha } \right)$, $
A_4 = 4\left( {\frac{y}{4} - \frac{1}{8}} \right)$ and $A_5 = \left( {\frac{{
5y}}{8} + \frac{{13}}{{16}}} \right)$. 

We skip all the technical details and refer to \cite{susy2} for them. The extended Hamiltonian turns out be
\begin{eqnarray}
H &=& \int {{d^3}x} \left[ {c\left( x \right){\partial _i}{\Pi ^i} - \frac{1}{2}{\Pi ^i}\frac{{\left( {{\nabla ^2} - {m^2}} \right)}}{{\left( {\alpha {\nabla ^2} - \beta } \right)}}{\Pi ^i} + \frac{1}{4}{F_{ij}}\frac{{\left( {{\nabla ^2} - {M^2}} \right)}}{{\left( {{\nabla ^2} - {m^2}} \right)}}{F^{ij}}} \right]    \nonumber\\
 &+& \int {{d^3}x} \frac{1}{2}{\partial ^i}{\partial ^k}{\Pi ^k}\frac{{\left( {{\nabla ^2} - {m^2}} \right)}}{{\left( {{\nabla ^2} + {\Omega ^2}} \right)\left( {\alpha {\nabla ^2} - \beta } \right)}}{\partial _i}{\partial _j}{\Pi _j}  \nonumber\\
 &+& \int {{d^3}} x\frac{{{A_5}}}{{{A_2}}}{t_{00}}\left( {\frac{{\left( {{\nabla ^2} - {m^2}} \right)}}{{\left( {\alpha {\nabla ^2} - \beta } \right)}}{\Pi _i} + \frac{{\left( {{\nabla ^2} - {m^2}} \right)}}{{\left( {{\nabla ^2} + {\Omega ^2}} \right)\left( {\alpha {\nabla ^2} - \beta } \right)}}} \right){\partial _i}{\partial _k}{\Pi _k}  \nonumber\\
&\times& \frac{{{\partial _i}{\partial _j}}}{{\left( {{\nabla ^2} - {m^2}} \right)}}\left( {\frac{{\left( {{\nabla ^2} - {m^2}} \right)}}{{\left( {\alpha {\nabla ^2} - \beta } \right)}}{\Pi ^j} + \frac{{\left( {{\nabla ^2} - {m^2}} \right)}}{{\left( {{\nabla ^2} + {\Omega ^2}} \right)\left( {\alpha {\nabla ^2} - \beta } \right)}}} \right){\partial ^j}{\partial ^m}{\Pi ^m},  \label{susy545}
\end{eqnarray}
to get the last line we used $c(x) = c_1 (x) - A_0 (x)$, $\alpha = \gamma -32t_{00} + 2\frac{{A_4 }}{{
A_2 }}$, $\beta = \gamma M^2 -32 t_{00} m^2$ and ${\textstyle{1 \over {{\Omega ^2}}}} = {\textstyle{{2{A_5}} \over {{A_2}}}}{t_{00}}{\textstyle{1 \over {\alpha {\nabla ^2} - \beta }}}$.

The corresponding expectation value, $\left\langle H \right\rangle _\Phi$, simplifies to
\begin{eqnarray}
\left\langle H \right\rangle _\Phi &=& \left\langle \Phi \right|\int {d^3 x} 
\Biggl[-\frac{1}{2}\Pi ^i \frac{{\left( {\nabla ^2 - m^2 } \right)}}{{\left( 
{\alpha \nabla ^2 - \beta } \right)}}\Pi _i \Biggr]|\Phi\rangle.
\label{susy550}
\end{eqnarray}

By following closely our previous work \cite{susy2}, we find that the static potential for two opposite charges located at $\mathbf{y}$ and $\mathbf{y^{\prime }}$ becomes:
\begin{eqnarray}
V = -\frac{{q^2 }}{{4\pi a }}\frac{{e^{ - \sqrt {{\raise0.5ex
\hbox{$\scriptstyle b $}\kern-0.1em/\kern-0.15em \lower0.25ex
\hbox{$\scriptstyle a $}}}\ L} }}{L}+ \frac{{q^2 m^2 }}{{8\pi a }}\ln \left( 
{1 + \frac{{\Lambda ^2 }}{{{\raise0.5ex\hbox{$\scriptstyle b $}\kern-0.1em/
\kern-0.15em\lower0.25ex\hbox{$\scriptstyle a $}}}}} \right)L,
\label{susy555}
\end{eqnarray}
where $\Lambda$ is an ultraviolet cutoff, $|\mathbf{y}-\mathbf{y}^{\prime }|\equiv L$, $a
= \gamma - 32{t_{00}} + 2\frac{{A_4}}{{A_2}}$ and $b = \gamma {M^2} - 32{
t_{00}}{m^2}$. To parallel the discussion of the preceding Subsection,
our interest here is  to find a meaning for the cutoff $\Lambda$. 
The first step in answering this question is to look at the 
definitions $A_1$, $A_2$, $A_3$, $A_4$, $a$, $b$, and $\gamma$.
Hence, we find that the only pole corresponding to a physical mass is 
the photino mass, previously given in eq. (\ref{susy475}). This then implies
that the previous static potential above makes sense only for distances above
the Compton wavelength of the photino, ${\lambda _{photino}} \equiv m_{photino}^{ - 1}$. 
One is thus lead to identify $\Lambda = {m_{photino}}$. It is important
to realize that whenever the pair particle-antiparticle is in static
interaction at a regime of distances $r > {\lambda _{photino}}$, the form of 
$V$ as given in eq. (\ref{susy555}) can be consistently taken. Accordingly,
the potential of Eq. (\ref{susy555}) takes the form 
\begin{equation}
V = -\frac{{q^2 }}{{4\pi a }}\frac{{e^{ - \sqrt {{\raise0.5ex
\hbox{$\scriptstyle b $}\kern-0.1em/\kern-0.15em \lower0.25ex
\hbox{$\scriptstyle a $}}}\ L} }}{L}+ \frac{{q^2 m^2 }}{{8\pi a }}\ln \left( 
{1 + \frac{{m_{photino}^2 }}{{{\raise0.5ex\hbox{$\scriptstyle b $}\kern
-0.1em/\kern-0.15em\lower0.25ex\hbox{$\scriptstyle a $}}}}} \right)L.
\label{susy560}
\end{equation}

\subsection{Supersymmetric extension of the Carroll-Field-Jackiw model for electrodynamics III}

As already expressed, in this Subsection we first carry out the dimensional reduction for a $(1+3)$-dimensional Lorentz violation Lagrangian in the matter and gauge sectors in a supersymmetric scenario. We shall focus our attention on a $(1+2)$- dimensional space-time, so that planar phenomena may be considered. Again, we will obtain an effective model for photons induced by the effects of SUSY in our framework with LIV.

We begin our discussion with the fermionic Lagrangian. We first observe that
in $(1+3)$-D the LV-Lagrangian for the matter field $\Psi \ $ reads
\begin{equation}
\mathcal{L}_f = \bar{\Psi} \Gamma^{\hat{\mu}}(i \partial_{\hat{\mu}} + \bar{a}_{\hat{\mu}} +\bar{b}_{\hat{\mu}}\Gamma_5 - m) \Psi, \label{susy565}
\end{equation}
where $\bar{b}^{\hat{\mu}} = \frac{1}{4}(W+V)^{\hat{\mu}} - b^{\hat{\mu}}$ , $\bar{a}^{\hat{\mu}} = \frac{1}{4}(W-V)^{\hat{\mu}} - a^{\hat{\mu}}$ , $W^{\hat{\mu}}(V) = \bar{\Lambda}_{+(-)}
\Gamma^{\hat{\mu}} \Gamma_5 \Lambda_{+(-)}$ are given in ref.\cite{susy3}  and represent the LV-backgroung parameters.  Besides, the bosonic part is given by
\begin{eqnarray}
\mathcal{L}_b = -\frac{1}{2} \phi_1^* \Big{(} \partial^{\hat{\mu}} \partial_{\hat{\mu}} + m^2 + i 2 \sqrt{2}(a+b)^{\hat{\mu}} \partial_{\hat{\mu}} \Big{)} \phi_1 -\frac{1}{2} \phi_2^* \Big{(}\partial^{\hat{\mu}} \partial_{\hat{\mu}} + m^2 + i 2 \sqrt{2}(a+b)^{\hat{\mu}} \partial_{\hat{\mu}} \Big{)} \phi_2. \label{susy570}
\end{eqnarray}

It should be further noted that beyond the quadratic terms, this method brings us a different kind of interaction between scalar and fermionic fields through the following interaction Lagrangian
\begin{equation}
\mathcal{L}_{int} = \Phi^\dagger v + h.c., \label{susy575}
\end{equation}
where $\Phi^\dagger = \begin{pmatrix}
\phi_1^* & \phi_2^*
\end{pmatrix}$ and $v = (i \bar{\Lambda} \Gamma \cdot\partial +\frac{m}{2} \bar{\Lambda}_R) \begin{pmatrix}\Psi\\  \Gamma_5 \Psi\end{pmatrix} $ . 

We can now to apply the dimensional reduction in the total lagrangian $\mathcal{L}_{tot} = \mathcal{L}_f + \mathcal{L}_b + \mathcal{L}_{int} $.

Following the conventions given in the Appendix of ref. \cite{susy3}, the corresponding dimensional reduction of the Lagrangian is carried out with $\Psi = \begin{pmatrix}\Psi_1 & \Psi_2 \end{pmatrix}^T$. We thus obtain
\begin{eqnarray}
\mathcal{L}_f \mspace{-6mu} &=&\mspace{-8mu}  \begin{pmatrix}\bar{\Psi}_1 & \bar{\Psi}_2\end{pmatrix} 
\mspace{-8mu} \begin{pmatrix}i \gamma \cdot \partial + \gamma\cdot\bar{a}- \bar{b}_3 + m  & - i \gamma \cdot \bar{b} + i \bar{a}_3 \\- i \gamma \cdot \bar{b} + i \bar{a}_3 & -i \gamma \cdot \partial - \gamma\cdot\bar{a}+ \bar{b}_3 + m\end{pmatrix} 
\mspace{-8mu}\begin{pmatrix}\Psi_1 \\ \Psi_2 \end{pmatrix}, \label{susy580}
\end{eqnarray}
where $\bar{\Psi}_1 = \Psi_1^{\dagger} \gamma^0 $. The bosonic and the mixed part can be evaluated together and they are given by
\begin{equation}
\mathcal{L}_{b+int} = \Phi^{\dagger} O(\partial) \Phi + \Phi^{\dagger} v + v^{\dagger} \Phi, \label{susy585}
\end{equation}
where, 
\begin{equation}
O(\partial) = \frac{1}{2} \Big(\partial^{\mu} \partial_{\mu} + m^2 + i 2 \sqrt{2}(a+b)^{\mu} \partial_{\mu}\Big), \label{susy590}
\end{equation} 
and 
\begin{eqnarray}
v&=&(i \bar{\Lambda}_1 \gamma \cdot\partial +\frac{m}{2} \bar{\Lambda}_{1R}) \begin{pmatrix}1 & 0\\0 & 1 \end{pmatrix} \begin{pmatrix}\Psi_1 \\ \Psi_2 \end{pmatrix} 
+ (i \bar{\Lambda}_2 \gamma \cdot\partial +\frac{m}{2} \bar{\Lambda}_{2R})\begin{pmatrix}0 & 1 \\ -1 & 0 \end{pmatrix} \begin{pmatrix}\Psi_1 \\ \Psi_2 \end{pmatrix} \nonumber\\
&=& \begin{pmatrix} \bar{Q}_1 & \bar{Q}_2 \\ - \bar{Q}_2 & \bar{Q}_1 \end{pmatrix} \begin{pmatrix}\Psi_1 \\ \Psi_2 \end{pmatrix}. \label{susy595}
\end{eqnarray} 

The $\bar{Q}_1$ and $\bar{Q}_2$ terms are given by $\bar{Q}_1=(i \bar{\Lambda}_1 \gamma \cdot\partial +\frac{m}{2} \bar{\Lambda}_{1R}) $ and $\bar{Q}_2 = (i \bar{\Lambda}_2 \gamma \cdot\partial +\frac{m}{2} \bar{\Lambda}_{2R})$. 

It is worth noting here that by making use of the shift in the field $\Phi \rightarrow \Phi + O(\partial)^{-1}v$, the previous equation can be written alternatively in the form 
\begin{equation}
\mathcal{L}_{b+ int} = \Phi^{\dagger} O(\partial) \Phi - v^{\dagger} O(\partial)^{-1} v, \label{susy600}
\end{equation} 
where 
\begin{eqnarray}
v^{\dagger} O(\partial)^{-1} v \mspace{-6mu}\,\, = \,\, \mspace{-8mu} \begin{pmatrix}\bar{\Psi}_1 & \bar{\Psi}_2\end{pmatrix} 
\,\, \mspace{-10mu}  \begin{pmatrix} Q_1 \bar{Q_1} - Q_2 \bar{Q_2} &Q_1 \bar{Q_2} + Q_2 \bar{Q_1}\\ -Q_1 \bar{Q_2} - Q_2 \bar{Q_1} & Q_2 \bar{Q_2} -  Q_1 \bar{Q_1} \end{pmatrix} 
\,\,\mspace{-8mu} O^{-1}(\partial) \begin{pmatrix}\Psi_1 \\ \Psi_2 \end{pmatrix},  \label{susy605}
\end{eqnarray}
and $O^{-1}(\partial) = \Big(\partial^{\mu} \partial_{\mu} + m^2 + i 2 \sqrt{2}(a+b)^{\mu} \partial_{\mu}\Big)^{-1}$. 

We thus find that the total Lagrangian becomes
\begin{eqnarray}
\mathcal{L}_{tot}\mspace{-6mu} &=& \mspace{-8mu} \begin{pmatrix}\bar{\Psi}_1 & \bar{\Psi}_2\end{pmatrix} 
\,\, \mspace{-8mu}  \begin{pmatrix}i \gamma \cdot \partial + \gamma\cdot\bar{a}- \bar{b}_3 + m  & - i \gamma \cdot \bar{b} + i \bar{a}_3 \\- i \gamma \cdot \bar{b} + i \bar{a}_3 & -i \gamma \cdot \partial - \gamma\cdot\bar{a}+ \bar{b}_3 + m\end{pmatrix}\
\mspace{-8mu} \begin{pmatrix}\Psi_1 \\ \Psi_2 \end{pmatrix}  \nonumber\\
&+&\mspace{-8mu}  \sum_{i=1}^2 \phi_i O(\partial) \phi_i + \begin{pmatrix}\bar{\Psi}_1 & \bar{\Psi}_2\end{pmatrix} \Omega(\partial, \Lambda) O^{-1}(\partial) \begin{pmatrix}\Psi_1 \\ \Psi_2 \end{pmatrix}, \nonumber\\
\label{susy610}
\end{eqnarray}
where $ \Omega(\partial, \Lambda) =  \begin{pmatrix} Q_1 \bar{Q_1} - Q_2 \bar{Q_2} &Q_1 \bar{Q_2} + Q_2 \bar{Q_1}\\ -Q_1 \bar{Q_2} - Q_2 \bar{Q_1} & Q_2 \bar{Q_2} -  Q_1 \bar{Q_1} \end{pmatrix}$. 

We now want to extend what we have done to the gauge sector.
 
By the introduction of a background scalar superfield 
\begin{eqnarray}
S = s + \sqrt2 \theta \chi + i\bar{\theta}\sigma^{\hat{\mu}} \theta \partial_{\hat{\mu}} s + \theta^2 F + \frac{i}{\sqrt2}\theta^2 \bar{\theta}\bar{\sigma}^{\hat{\mu}} \partial_{\hat{\mu}}\chi 
- \frac{1}{4} \bar{\theta}^2 \theta^2 \Delta \ s, \label{susy615}
\end{eqnarray}
with the properties $(s + s^*) = 0$, $ (s - s^*) = - \frac{i}{2} x_{\hat{\mu}} \xi^{\hat{\mu}}$ and $\partial_{\hat{\mu}} \chi = 0$, we are able to write a LV-action. These properties have a meaning that the SUSY breaks down and generates a non-null vector background $\xi$ and a non-null fermionic parameter $\chi$. The action is given by:
\begin{equation}
S_{CPT-odd} = \int d^4 x d^4 \theta \Big{(} W^\alpha (D_\alpha V) S + W^{\dot{\alpha}} (D_{\dot{\alpha}} V) S \Big{)}, \label{susy620}
\end{equation}
where $V$ is the Vector superfield in the Wess-Zumino gauge and $W^\alpha = -\frac{1}{4}(\bar{D})^2D^{\alpha} V $. Rewriting in terms of the component fields the total Lagrangian is written as, $\mathcal{L}_{tot-gauge} = \mathcal{L}_{A} + \mathcal{L}_{ph} + \mathcal{L}_{int-gauge}$, where 
\begin{equation}
\mathcal{L}_A = -\frac{1}{4}{F_{\hat{\mu} \hat{\nu}} }^2 + \frac{1}{2} \epsilon^{\hat{\mu} \hat{\nu} \hat{\alpha} \hat{\beta}} \xi_{\hat{\mu}} A_{\hat{\nu}} F_{\hat{\alpha} \hat{\beta}}. \label{susy625}
\end{equation}
For the photino ($\lambda$) we have (with fermionic Lorentz breaking parameter ($\chi$)
\begin{eqnarray}
\mathcal{L}_{ph} \!\!\!\!&=&\!\!\!\! - \frac{i}{2} \bar{\lambda}\Gamma^{\hat{\mu}}\partial_{\hat{\mu}} \lambda + \phi \bar{\lambda} \lambda 
-i \rho \bar{\lambda} \Gamma_5 \lambda - \bar{V}_{\hat{\mu}} \bar{\lambda} \Gamma^{\hat{\mu}}\Gamma_5 \lambda, \label{susy630}
\end{eqnarray}
with $\phi = \Big[Re(F)+ \frac{1}{4}\bar{\chi} \chi \Big]$ , $\rho = \Big[Im(F) + \frac{i}{4}\bar{\chi}\Gamma_5 \chi \Big]$ and $\bar{V}_{\hat{\mu}} = \frac{1}{4}\Big[V_{\hat{\mu}}+ \bar{\chi}\Gamma_{\hat{\mu}}\Gamma_5 \chi \Big]$. This procedure also given a new interaction term between the photon and the photino field, and this term is given by
\begin{equation}
\mathcal{L}_{int-gauge} = \sqrt{2} \bar{\lambda}\Gamma^{\hat{\mu} \hat{\nu}} \Gamma_5 \chi F_{\hat{\mu} \hat{\nu}}. \label{susy635}
\end{equation}

Now, we apply the dimensional reduction to the Lagrangean $\mathcal{L}_{tot-gauge} = \mathcal{L}_A + \mathcal{L}_{ph} + \mathcal{L}_{int-gauge}$.

From the dimensional reduction scheme of the Appendix of ref. \cite {susy3}, we write $A^{\hat{\mu}} = (A^\mu, \varphi)$ , 
$\lambda = \begin{pmatrix}\lambda_1 & \lambda_2 \end{pmatrix}^T$ and $\chi = \begin{pmatrix}\chi_1  & \chi_2 \end{pmatrix}^T$, we have
\begin{eqnarray}
\mathcal{L}_A = -\frac{1}{4} F_{\mu \nu}^2 + \frac{1}{2}\partial_\mu \varphi \partial^\mu \varphi  -\frac{\varphi}{4}\varepsilon^{\mu \nu \alpha} \xi_\mu \partial_\nu A_\alpha 
-\frac{1}{2}\xi_3 \varepsilon^{\mu \nu \alpha}A_\mu \partial_\nu A_\alpha. \label{susy640}
\end{eqnarray}
The photino sector is given by
\begin{eqnarray}
\mathcal{L}_{ph} \mspace{-6mu} \,\,= \,\, \mspace{-8mu}\begin{pmatrix}\bar{\lambda}_1 & \bar{\lambda}_2 \end{pmatrix}\begin{pmatrix}-\frac{i}{2} \gamma \cdot \partial +  \phi - \bar{V}_3 &  \rho - i \gamma \cdot \bar{V} \\ -\rho - i \gamma \cdot \bar{V} & \frac{i}{2} \gamma \cdot \partial + \phi + \bar{V}_3 \end{pmatrix} 
\,\,\,\mspace{-8mu}\begin{pmatrix}\lambda_1 \\ \lambda_2 \end{pmatrix}, \label{susy645}
\end{eqnarray}
where $\bar{V}_{\mu} = \frac{1}{4}\Big[V_{\mu}+ \bar{\chi}\Gamma_{\mu}\Gamma_5 \chi \Big]$ and $\bar{V}_{\hat{\mu}} = \frac{1}{4}\Big[V_{\hat{\mu}}+ \bar{\chi}\Gamma_{\hat{\mu}}\Gamma_5 \chi \Big]$. The mixing terms can be rewritten as follows:
\begin{eqnarray}
\mathcal{L}_{int-gauge} = \sqrt{2} i (\bar{\lambda}_1\gamma^{\mu \nu} \chi_2 + \bar{\lambda}_2\gamma^{\mu \nu} \chi_1)F_{\mu \nu} 
+ \sqrt{2}(\bar{\lambda}_1\gamma^{\mu} \chi_2 + \bar{\lambda}_2\gamma^{\mu} \chi_1) \partial_\mu \varphi. \label{susy650}
\end{eqnarray}
In a short way, we can rewrite the above equation as $\mathcal{L}_{int} = \bar{\Theta} \Upsilon $ , where $\bar{\Theta} = \begin{pmatrix}\bar{\lambda}_1 & \bar{\lambda}_2 \end{pmatrix}$ and
\begin{equation}
\Upsilon =\sqrt{2} \begin{pmatrix}\gamma \cdot F & \gamma \cdot \partial \varphi \\ \gamma \cdot \partial \varphi & \gamma \cdot F \end{pmatrix} \begin{pmatrix} \chi_2 \\ \chi_1 \end{pmatrix}. \label{susy655}
\end{equation}
Here $\gamma \cdot F = \gamma^{\mu \nu} F_{\mu \nu}$ and $\gamma \cdot \partial \varphi = \gamma^\mu \partial_\mu \varphi$. Manipulating the equation above and applying the same kind of shift used in the matter sector, we can rewrite the photino and mixing terms as
\begin{eqnarray}
\mathcal{L}_{ph+ int} = \bar{\Theta} O(\partial) \Theta + \bar{\Theta} \Upsilon = \bar{\Theta} O(\partial)\Big(\Theta +\frac{1}{2} O^{-1}(\partial) \Upsilon \Big) 
+ \frac{1}{2} \bar{\Theta} \Upsilon, \label{susy660}
\end{eqnarray}
where $\tilde{O}(\partial) = \begin{pmatrix}-\frac{i}{2} \gamma \cdot \partial +  \phi - \bar{V}_3 &  \rho - i \gamma \cdot \bar{V} \\ -\rho - i \gamma \cdot \bar{V} & \frac{i}{2} \gamma \cdot \partial + \phi + \bar{V}_3 \end{pmatrix}$. 

With the shift $\Theta \rightarrow \Theta + \frac{1}{2} \tilde{O}^{-1}(\partial) \Upsilon $ we have, finally
\begin{equation}
\mathcal{L}_{ph+ int} = \bar{\Theta} \tilde{O}(\partial) \Theta - \frac{1}{4} \bar{\Upsilon} \tilde{O}^{-1} \Upsilon. \label{susy665}
\end{equation}
Thus, the final action of the CPT-odd gauge sector will be given by
\begin{eqnarray}
 \mathcal{L}_{tot-gauge}\!\!\!\! &=&\!\!\!\! -\frac{1}{4} F_{\mu \nu}^2 + \frac{1}{2}\partial_\mu \varphi \partial^\mu \varphi  -\frac{\varphi}{4}\varepsilon^{\mu \nu \alpha} \xi_\mu \partial_\nu A_\alpha  
 -\frac{1}{2}\xi_3 \varepsilon^{\mu \nu \alpha}A_\mu \partial_\nu A_\alpha  
 - \frac{1}{4} \bar{\Upsilon} \tilde O^{-1} \Upsilon \nonumber\\
 &+& \!\!\!\! \begin{pmatrix}\bar{\lambda}_1 & \bar{\lambda}_2 \end{pmatrix} \tilde{O}(\partial) \begin{pmatrix}\lambda_1 \\ \lambda_2 \end{pmatrix}.  \label{susy670}
 \end{eqnarray}

In summary then, we have obtained a complete $(2+1)$-dimensional Lagrangian, which defines a new electrodynamics. In the following Section we compute the interaction energy between static point-like sources for this new electrodynamics. Following the same steps as the ones presented in our previous works \cite{susy1,susy2}, we get an effective photonics-scalar Lagrangian which shall be the matter in the coming Section.

We now pass on to the calculation of the interaction energy for the model under consideration,
following our earlier line of argument. We start off our analysis by considering the effective Lagrangian density
\begin{eqnarray}
{\cal L} &=&  - \frac{1}{4}F_{\mu \nu }^2 + \frac{m}{4}{\varepsilon ^{\mu \nu \kappa }}{A_\mu }{F_{\nu \kappa }} + m{\varepsilon ^{\mu \nu \kappa }}{v_\mu }{F_{\nu \kappa }}\varphi  
+ \frac{1}{2}{\left( {{\partial _\mu }\varphi } \right)^2} + {t_{\mu \nu }}{F^{\mu \lambda }}F_\lambda ^\nu  + \alpha {t_{\mu \nu }}{F^{\mu \lambda }}\frac{\Delta }{{\bar \Delta }}F_\lambda ^\nu 
 \notag \\
&+& \beta {t_{\rho \lambda }}{F^{\mu \lambda }}\frac{{{\partial _\mu }{\partial _\nu }}}{{\bar \Delta }}{F^{\nu \rho }} + {s^\mu }{F_{\mu \nu }}{\partial ^\nu }\varphi  + {s_\lambda }{F^{\mu \lambda }}\frac{{{\partial _\mu }\Delta }}{{\bar \Delta }}\varphi, \label{susy675}
\end{eqnarray}
where $\Delta \equiv \partial _\mu \partial ^\mu$, ${v_\mu } \equiv {\xi _\mu }$, ${t_{\mu \nu }}$ and $s_{\mu}$ are given in Ref. \cite{susy2}. By a further integration over the $\varphi$-field, we find that the previous effective Lagrangian density can be brought to the form:
\begin{eqnarray}
{\cal L}\mspace{-6mu} &=&\mspace{-8mu} -\frac{1}{4}{F_{\mu \nu }}\left( {1 - \frac{{4{m^2}{v^2}}}{\Delta }} \right){F^{\mu \nu }} + \frac{m}{4}{\varepsilon ^{\mu \nu \kappa }}{A_\mu }{F_{\nu \kappa }} 
+\mspace{-8mu}\,\, {t_{\mu \nu }}{F^{\mu \lambda }}F_\lambda ^\nu  + \alpha {t_{\mu \nu }}{F^{\mu \lambda }}\frac{\Delta }{{\bar \Delta }}F_\lambda ^\nu  + 
\beta {t_{\rho \lambda }}{F^{\mu \lambda }}\frac{{{\partial _\mu }{\partial _\nu }}}{{\bar \Delta }}{F^{\nu \rho }}  \notag \\
&+&\mspace{-8mu} 2{v_\mu }{v_\nu }{F^{\mu \lambda }}\frac{{{m^2}}}{\Delta }F_\lambda ^\nu  - m{\varepsilon _{\rho \xi \sigma }}{v^\xi }{s_\lambda }{F^{\mu \lambda }}\left( {\frac{1}{{\bar \Delta }} - \frac{1}{\Delta }} \right){\partial _\mu }{F^{\rho \sigma }} 
+\mspace{-8mu}\,\, \frac{1}{2}{s_\lambda }{s_\nu }{F^{\mu \lambda }}\left( {\frac{\Delta }{{{{\left( {\bar \Delta } \right)}^2}}} - \frac{2}{{\bar \Delta }} + \frac{1}{\Delta }} \right){\partial _\mu }{\partial _\rho }{F^{\nu \rho }}. \nonumber\\
\label{susy680}
\end{eqnarray}
As we have noted before, by studying the static potential one may replace $\Delta$ by $-\nabla ^2
$ in Eq.(\ref{susy680}). Let us also recall here that the only
non-vanishing $t_{\mu \nu }$-terms are the diagonal ones, since 
$t_{\mu \nu }$ can be brought into a diagonal form. 
We further note that, without loss of generality, we may always choose $t_{00}\ne 0$.
By considering the ${v^i} \ne 0$ and ${v^{ij}} = 0$ ($v_{0}=0$) case (referred to as the space-like background in what follows), the following effective Lagrangian can be studied
\begin{eqnarray}
{\cal L}\mspace{-6mu} &=&\mspace{-8mu} - \frac{1}{4}{F_{\mu \nu }}\left( {1 - \frac{{4{m^2}{{\bf v}^2}}}{{{\nabla ^2}}}} \right){F^{\mu \nu }} + \frac{m}{4}{\varepsilon ^{\mu \nu \kappa }}{A_\mu }{F_{\nu \kappa }} 
+\mspace{-8mu}\, \,\,{t_{00}}\frac{{\left( {{A_2} + \alpha } \right)}}{{{A_2}}}{F^{i0}}{\cal O}{F^{i0}}  
 - \beta \frac{{{t_{00}}}}{{{A_2}}}{F^{i0}}\frac{{{\partial _i}{\partial _j}}}{{\left( {{\nabla ^2} - {X^2}} \right)}}{F^{j0}} \notag \\
&+&\mspace{-8mu} \frac{{2m}}{{{A_2}}}\left( {{\bf v} \cdot {\bf s}} \right){F^{j0}}{{\cal O}^ \prime }{\partial _j}B + B{{\cal O}^{ \prime  \prime }}B -A_{0}J^{0}, \label{susy685}
\end{eqnarray}
where $B$ is the magnetic field ($B = {\varepsilon _{ij}}{\partial ^i}{A^j}$), ${A_1} = {\mu ^2}$ and ${A_2} \equiv \left( {{c^{ii}} - 1} \right) = \left( {k - 1} \right)$. Notice that these $A_{1}$ and $A_{2}$ are not to be confused with the components of the photon field. Nevertheless ${\cal O} \equiv \left[ {\frac{{{\nabla ^4} - p{\nabla ^2} - q}}{{{\nabla ^2}\left( {{\nabla ^2} - {X^2}} \right)}}} \right]$, 
${{\cal O}^ \prime } \equiv \left[ {\frac{{\left( {1 - {A_2}} \right){\nabla ^2} + {A_1}}}{{{\nabla ^2}\left( {{\nabla ^2} - {X^2}} \right)}}} \right]$ and ${{\cal O}^{ \prime  \prime }} \equiv \left[ {\frac{{- {\bar v}{\nabla ^2} + w}}{{{\nabla ^2}\left( {{\nabla ^2} - {X^2}} \right)}}} \right]$. Here $p = \frac{{\left( {{t_{00}}{A_1} - 2{m^2}{{\bf v}^2}{A_2}} \right)}}{{{t_{00}}\left( {{A_2} + \alpha } \right)}}$, $q = \frac{{\left( {2{m^2}{{\bf v}^2}{A_1}} \right)}}{{{t_{00}}\left( {{A_2} + \alpha } \right)}}$, ${X^2} = \frac{{{A_1}}}{{{A_2}}}$, $\bar v = 4{m^2}{{\bf v}^2}$ and $w = 4{m^2}{{\bf v}^2}{X^2}$.

As was explained in \cite{susy3}, the Hamiltonian is given by   
\begin{eqnarray}
{H} \mspace{-6mu}&=&\mspace{-8mu} \int {{d^2}x} \left\{ { c(x)\left( {{\partial _i}{\Pi ^i} + \frac{m}{2}{\varepsilon ^{ij}}{\partial _i}{A_j} - {J^0} + \frac{1}{2}{E_i}\Lambda {D_{ij}}{E_j} + \frac{{2m}}{{{A_2}}}\left( {{\bf v} \cdot {\bf s}} \right){E^i}{O^\prime}{\partial _i}B  } \right)} \right\} \nonumber\\
 &+&\mspace{-8mu} \int {{d^2}x} \left\{ {\frac{1}{2}B\left( {1 - \frac{{4{m^2}{v^2}}}{{{\nabla ^2}}} - 2{O^{\prime \prime}}} \right)B}\right\}, \label{susy690}
\end{eqnarray}
where $c(x) = c_1 (x) - A_0 (x)$. 

In order to illustrate the discussion, we now write the Dirac brackets (\ref{susy65})  in terms of the magnetic ($B = {\varepsilon _{ij}}{\partial ^i}{A^j}$) and electric 
(${E_i}\mspace{-6mu}\,\, = \, \,\mspace{-6mu}{\Lambda ^{ - 1}}\left( {{\delta _{ij}} + \frac{{{\partial _i}{\partial _j}}}{{\left( {{\gamma ^2}\Lambda  - {\nabla ^2}} \right)}}} \right) 
\mspace{-6mu}\left( {{\Pi _j} + \frac{{4m}}{{{A_2}}}\left( {{\bf v} \cdot {\bf s}} \right){\varepsilon _{kl}}{\partial ^k}{\partial _j}{A^l} - \frac{m}{2}{\varepsilon _{jk}}{A^k}} \right)$) fields as
\begin{eqnarray}
{\left\{ {{E_i}\left( {\bf x} \right),{E_r}\left( {\bf y} \right)} \right\}^ * } \mspace{-9mu}&=&\mspace{-9mu} \frac{{2m}}{{{A_2}}}{\Lambda ^{ - 2}}\left( {{\bf v} \cdot {\bf s}} \right) {{\cal O}^{\prime}} \left( {{\varepsilon _{kr}}{\partial ^k}{\partial _i} - {\varepsilon _{pi}}{\partial ^p}{\partial _r}} \right) 
\left( {1 + \frac{{{\nabla ^2}}}{\Omega }} \right){\delta ^{\left( 2 \right)}}\left( {{\bf x} - {\bf y}} \right) \nonumber\\
&+& m{\Lambda ^{ - 2}}D_{ij}^{ - 1}D_{rn}^{ - 1}{\varepsilon _{nj}}{\delta ^{\left( 2 \right)}}\left( {{\bf x} - {\bf y}} \right), \label{susy695}
\end{eqnarray}
where $D_{ij}^{ - 1} = {\delta _{ij}} + \frac{{{\partial _i}{\partial _j}}}{{\left( {\Lambda {\gamma ^2} - {\nabla ^2}} \right)}}$, $ 1 + \frac{{{\nabla ^2}}}{\Omega } = 1 + 2\beta {t_{00}}\frac{{{\nabla ^2}}}{{{A_2}{\gamma ^2}\left( {{\nabla ^2} - {X^2}} \right) - 2\beta {t_{00}}{\nabla ^2}}}$ and ${\gamma ^2} = \frac{{{A_2}}}{{2\beta {t_{00}}}}\left( {{\nabla ^2} - {X^2}} \right)$. 

And,
\begin{equation}
{\left\{ {B\left( {\bf x} \right),B\left( {\bf y} \right)} \right\}^ * } = 0, \label{susy670}
\end{equation}
\begin{eqnarray}
{\left\{ {{E_i}\left( {\bf x} \right),B\left( {\bf y} \right)} \right\}^ * } =  - {\Lambda ^{ - 1}}{\varepsilon _{ij}}{\partial _j}{\delta ^{\left( 2 \right)}}\left( {{\bf x} - {\bf y}} \right). \label{susy675}
\end{eqnarray}

One can now easily derive the equations of motion for the magnetic and electric fields. We find
\begin{equation}
\dot B\left( {\bf x} \right) =  - {\varepsilon _{ij}}{\partial _i}{E_j}\left( {\bf x} \right), \label{susy680}
\end{equation}
and
\begin{eqnarray}
{\dot E_i}\left( {\bf x} \right) &=& 
\frac{{2m}}{{{A_2}}}{\Lambda ^{ - 1}}\left( {v \cdot s} \right) {{\cal O}^\prime} \left( {{\varepsilon _{kr}}{\partial _i} - {\varepsilon _{ki}}{\partial _r}} \right)\left( {1 + \frac{{{\nabla ^2}}}{\Omega }} \right) 
{D_{rb}}{\partial ^k}{E_b}\left( {\bf x} \right) \nonumber\\
&+& {\Lambda ^{ - 1}}{\varepsilon _{ij}}\left( {1 - \frac{{4{m^2}{{\bf v}^2}}}{{{\nabla ^2}}} - 2{{\cal O}^{ \prime  \prime }}} \right){\partial _j}{B}\left( x \right)     
+ \frac{{2m}}{{{A_2}}}{\Lambda ^{ - 1}}{\varepsilon _{ij}}{\partial _j}{\partial _k}{{\cal O}^ \prime }{E_k}\left( {\bf x} \right). \label{susy685}
\end{eqnarray}
In the same way, we write the Gauss law as
\begin{equation}
{D_{ij}}\Lambda {\partial _i}{E_j} - mB - \frac{{4m}}{{{A_2}}}\left( {{\bf v} \cdot {\bf s}} \right){{\cal O}^ \prime }{\nabla ^2}B = {J^0},\label{susy690}
\end{equation}
where ${D_{ij}} = \left( {{\delta _{ij}} - \frac{{{\partial _i}{\partial _j}}}{{\Lambda {\gamma ^2}}}} \right)$.

This implies that for static fields, equations (\ref{susy680}) and (\ref{susy685}) must vanish. The magnetic field then reads 
\begin{equation}
B =  - \frac{{m\left[ {\left( {{\bf v} \cdot {\bf s}} \right) - 1} \right]}}{{{A_2}}}\frac{{\left[ {\left( {1 - {A_2}} \right){\nabla ^2} + {A_1}} \right]}}{{\left( {{\nabla ^2} - {X^2}} \right)\left( {{\nabla ^2} + 4{m^2}{{\bf v}^2}} \right)}}{\partial _i}{E_i}. \label{susy695}
\end{equation}
With the aid of equations (\ref{susy695}) and (\ref{susy690}) we find 
\begin{eqnarray}
{E_i}({\bf x})\mspace{-8mu} &=&\mspace{-8mu} \frac{1}{{{g_1}}}{\partial _i}\left\{ {\frac{{\left[ {{\nabla ^4} + \left( {4{m^2}{{\bf v}^2}{X^2}} \right){\nabla ^2} - 4{m^2}{{\bf v}^2}{X^2}} \right]}}{{{\nabla ^2}\left[ {{\nabla ^4} + \frac{{{g_2}}}{{{g_1}}}{\nabla ^2} + \frac{{{g_3}}}{{{g_1}}}} \right]}}} \right\} 
\mspace{-8mu}\left( { - {J^0}} \right). \label{susy700}
\end{eqnarray}
Here we have simplified our notation by setting ${g_1} = 1 + 2{t_{00}} + \frac{{2{t_{00}}\left( {\alpha  - \beta } \right)}}{{{A_2}}}$, ${g_2} = \left( {1 + 2{t_{00}}} \right){X^2} + \left\{ {\left[ {1 - \left( {{\bf v} \cdot {\bf s}} \right)} \right]\frac{{\left( {1 - {A_2}} \right)}}{{{A_2}}} - 4{{\bf v}^2}\left( {1 + 2{t_{00}} + 2{t_{00}}\frac{{\left( {\alpha  - \beta } \right)}}{{{A_2}}}} \right)} \right\}$ and ${g_3} = {m^2}{X^2}\left[ {\left( {{\bf v} \cdot {\bf s}} \right) - 4{{\bf v}^2}\left( {1 + 2{t_{00}}} \right) - 1} \right]$. From equation (\ref{susy700}) it follows now, after some manipulations, that     
\begin{eqnarray}
{E_i}\left( {\bf x} \right) \mspace{-6mu} &=& -\mspace{-9mu} \frac{1}{{{g_1}}}\frac{1}{{\left( {M_1^2 - M_2^2} \right)}}{\partial _i}\left[ {\frac{{{\nabla ^2}}}{{\left( {{\nabla ^2} - M_1^2} \right)}} - \frac{{{\nabla ^2}}}{{\left( {{\nabla ^2} - M_2^2} \right)}}} \right] 
\mspace{-9mu} \left( { - {J^0}} \right)   \nonumber\\
 &+&\mspace{-9mu}\frac{1}{{{g_1}}}\frac{{\left( {{X^2} - 4{m^2}{{\bf v}^2}} \right)}}{{\left( {M_1^2 - M_2^2} \right)}}{\partial _i}\left[ {\frac{1}{{\left( {{\nabla ^2} - M_1^2} \right)}} - \frac{1}{{\left( {{\nabla ^2} - M_2^2} \right)}}} \right] 
 \mspace{-9mu}\left( { - {J^0}} \right)  \nonumber\\
&+&\mspace{-9mu}\frac{{\left( {4{m^2}{v^2}{X^2}} \right)}}{{{g_1}\left( {M_1^2 - M_2^2} \right)}}\frac{{{\partial _i}}}{{{\nabla ^2}}}\left[ {\frac{1}{{\left( {{\nabla ^2} - M_1^2} \right)}} - \frac{1}{{\left( {{\nabla ^2} - M_2^2} \right)}}} \right] 
\mspace{-9mu}\left( { - {J^0}} \right), \label{susy705} 
 \end{eqnarray}
where $M_1^2 =  - \frac{1}{2}\frac{{{g_2}}}{{{g_1}}} + \frac{1}{2}\sqrt {\frac{{g_2^2}}{{g_1^2}} - 4\frac{{{g_3}}}{{{g_1}}}}$ and $M_2^2 =  - \frac{1}{2}\frac{{{g_2}}}{{{g_1}}} - \frac{1}{2}\sqrt {\frac{{g_2^2}}{{g_1^2}} - 4\frac{{{g_3}}}{{{g_1}}}}$.
For ${J^0}\left( {\bf x} \right) = q{\delta ^{\left( 2 \right)}}\left( {\bf x} \right)$, expression (\ref{susy705}), becomes
\begin{eqnarray}
{E_i}\left( {\bf x} \right) &=&  - \frac{q}{{{g_1}}}\frac{1}{{\left( {M_1^2 - M_2^2} \right)}}{\partial _i}\left\{ {{\nabla ^2}{G_1}\left( {\bf x} \right) - {\nabla ^2}{G_2}\left( {\bf x} \right)} \right\} 
+ \frac{q}{{{g_1}}}\frac{{\left( {{X^2} - 4{m^2}{{\bf v}^2}} \right)}}{{\left( {M_1^2 - M_2^2} \right)}}{\partial _i}\left\{ {{G_1}\left( {\bf x} \right) - {G_2}\left( {\bf x} \right)} \right\}    \nonumber\\
&+&\frac{q}{{{g_1}}}\frac{{\left( {4{m^2}{{\bf v}^2}{X^2}} \right)}}{{\left( {M_1^2 - M_2^2} \right)}}{\partial _i}\left\{ {\frac{{{G_1}\left( {\bf x} \right)}}{{{\nabla ^2}}} - \frac{{{G_1}\left( {\bf x} \right)}}{{{\nabla ^2}}}} \right\},\label{susy710}
\end{eqnarray}
where ${G_1}\left( {\bf x} \right) =  - \frac{{{\delta ^{\left( 2 \right)}}\left( {\bf x} \right)}}{{{\nabla ^2} - M_1^2}} = \frac{1}{{2\pi }}{K_0}\left( {{M_1}|{\bf x}|} \right)$ and ${G_2}\left( {\bf x} \right) =  - \frac{{{\delta ^{\left( 2 \right)}}\left( {\bf x} \right)}}{{{\nabla ^2} - M_2^2}} = \frac{1}{{2\pi }}{K_0}\left( {{M_2}|{\bf x}|} \right)$.

Inserting the expression (\ref{susy710}) in equation (\ref{susy115}), we obtain
\begin{eqnarray}
{{\cal A}_0}({\bf x}) &=& - \frac{q}{{{g_1}}}\frac{1}{{\left( {M_1^2 - M_2^2} \right)}}\left( {{\nabla ^2}{G_1}\left( {\bf x} \right) - {\nabla ^2}{G_2}\left( {\bf x} \right)} \right) 
+ \frac{q}{{{g_1}}}\frac{{\left( {{X^2} - 4{m^2}{{\bf v}^2}} \right)}}{{\left( {M_1^2 - M_2^2} \right)}}\left( {{G_1}\left( {\bf x} \right) - {G_2}\left( {\bf x} \right)} \right) \nonumber\\
&+& \frac{q}{{{g_1}}}\frac{{\left( {4{m^2}{{\bf v}^2}{X^2}} \right)}}{{\left( {M_1^2 - M_2^2} \right)}}\left( {\frac{{{G_1}\left( {\bf x} \right)}}{{{\nabla ^2}}} - \frac{{{G_2}\left( {\bf x} \right)}}{{{\nabla ^2}}}} \right), \label{susy715}
\end{eqnarray}
after subtracting the self-energy terms.

And, finally, with the aid of relations (\ref{susy110}) and (\ref{susy715}), it follows that the potential for two opposite charges, located at ${\bf 0}$ and ${\bf y}$, is given by
\begin{eqnarray}
V &=&  - \frac{{{q^2}}}{{2\pi {g_1}}}\frac{{\left( {{X^2} - 4{m^2}{{\bf v}^2}} \right)}}{{\left( {M_1^2 - M_2^2} \right)}}\left( {{K_0}\left( {{M_1}L} \right) - {K_0}\left( {{M_2}L} \right)} \right) 
+\frac{{{q^2}}}{{{g_1}}}\frac{{{m^2}{{\bf v}^2}{X^2}}}{{{M_1}{M_2}\left( {{M_1} + {M_2}} \right)}}\,\,L    \nonumber\\
&+& \frac{{{q^2}}}{{2\pi {g_1}}}\frac{1}{{\left( {M_1^2 - M_2^2} \right)}}\left( {{\nabla ^2}{K_0}\left( {{M_1}L} \right) - {\nabla ^2}{K_0}\left( {{M_2}L} \right)} \right),  
\label{susy720}
\end{eqnarray}
where $L \equiv  |{\bf y}|$. In this last line, we have used that $\frac{{{G_1}\left( {\bf x} \right)}}{{{\nabla ^2}}} = \frac{{|{\bf x}|}}{{4{M_1}}}$ and $\frac{{{G_2}\left( {\bf x} \right)}}{{{\nabla ^2}}} = \frac{{|{\bf x}|}}{{4{M_2}}}$.

\section{Discussion and Conclusions}

This contribution sets out to present a particular scenario where Lorentz-symmetry violating effects emerge in connection with supersymmetry. As we have already pointed out and cited a number of reference papers in the Introduction, there are different approaches to treat both LSV and SUSY in a single framework. Ours consists in adopting the standpoint that the background responsible for the breaking of Lorentz symmetry has its microscopic origin traced back to some supersymmetric multiplet, so that, besides the bosonic (non-dynamical) fields that respond for the Lorentz violation, there appears a background fermion sector, whose parity-preserving (and eventual parity-breaking condensates) directly affect the mass spectrum of the gauge sector and promotes de photon-photino mass splitting, as derived in the Subsections 2.2 and 2.3. We have actually concluded that the condensate of the background fermion induces a gaugino mass shift, as described by eq. (\ref{susy355b}). We however recall that the photon is massive even in the non-supersymmetric extension of the CFJ model.

Another aspect we would like to highlight concerns the possibility to formulate effective photonic models by integrating over the photino degrees of freedom.  According to our procedure developed in Subsections 2.2 and 2.3, we reshuffle the photino field by a suitable shift in fiel space and we are able thereby to eliminate the mixing between the photon and the photino fields induced by the presence of the background fermion. Doing so, we decouple the photino field and by integrating out its degrees of freedom we get effective photonic models that carry information about Lorentz violation and exhibit vacuum birrefringence and dichroism. 
 
Besides, in the Subsection 2.4, we have carried the dimensional reduction of the supersymmetric version of  the Carroll-Field-Jackiw model to $(1+2)$ dimensions to set up a framework of possible discussions in connection with lower-dimensional Condensed-Matter Physics phenomena. In all models we have contemplated throughout the present contribution, we have calculated and discussed the interparticle potentials that follow from the Lorentz-symmetry violating effective photonic models worked out upon integration of the photino degrees of freedom.

Finally, we have also studied the confinement versus screening issue for our effective photonics models, giving a specific significance for the fermion condensates that characterize the background responsible for the LSV. As we have shown the impact of these condensates become manifest in the interaction energy for these effective models. It should be emphasized that to illustrate these interesting consequences we have exploited the gauge-invariant but path-dependent formalism. We notice that an important feature of this formalism is a correct identification of physical degrees of freedom for understanding the physics hidden in gauge theories. We thus find that the interaction energy contains a linear confining term and a Yukawa type potential in $(1+3)$ dimensions. Although in $(1+2)$ dimensions, we obtain a screening part, encoded by Bessel functions, and a linear confining potential.
 
We conclude this contribution by saying that our major effort here has been to bring together a number of results we have attained over the past eight years in trying to investigate a possible particular scenario that connects Lorentz-symmetry violating effects with supersymmetry by attributing to the latter the microscopic origin of the background entities that characterize the so-called $k_{AF}$- and $k_{F}$-models which are particular representatives of the general category of models referred to as the Standard Model Extension. As an immediate follow-up, we wish to pursue a calculation of the Primakoff-like conversion of massive photini into photons. We shall, in practice, derive the decay rate (inverse lifetime), $\Gamma_{photino \to \gamma}$ , of a photino to convert to a photon. We are going to report on that elsewhere.

\vspace{6pt} 





\section{acknowledgements}

One of us (P. G.) was partially supported by Fondecyt (Chile) grant 1180178 and by ANID PIA / APOYO AFB180002.

\end{document}